\documentclass[12 pt]{article}
\usepackage{graphicx}
\usepackage{harpoon}
\usepackage{mathtools}
\usepackage{chemarrow}
\usepackage{color}
 \usepackage{stackrel}
\usepackage{latexsym}
\usepackage{amssymb}
\usepackage{amsmath}
\usepackage{mdframed}
\usepackage{hyperref, url}
\usepackage[show]{ed}
\usepackage{subfigure}
\usepackage[ntheorem]{empheq}
\usepackage{amsthm, amssymb, amsfonts, latexsym}
\mathchardef\mhyphen="2D

\usepackage{pst-node}

\makeatletter
\DeclareRobustCommand{\cev}[1]{%
  \mathpalette\do@cev{#1}%
}
\newcommand{\do@cev}[2]{%
  \fix@cev{#1}{+}%
  \reflectbox{$\m@th#1\vec{\reflectbox{$\fix@cev{#1}{-}\m@th#1#2\fix@cev{#1}{+}$}}$}%
  \fix@cev{#1}{-}%
}
\newcommand{\fix@cev}[2]{%
  \ifx#1\displaystyle
    \mkern#20mu
  \else
    \ifx#1\textstyle
      \mkern#20mu
    \else
      \ifx#1\scriptstyle
        \mkern#26mu
      \else
        \mkern#26mu
      \fi
    \fi
  \fi
}

\makeatother

\newcommand{\fig}[1]{\epsfbox}
\newcommand{\bp}{\begin{minipage}{3.1cm}}
\newcommand{\ep}{\end{minipage}}
\usepackage{pbox}

\newtheorem{Theorem}{Theorem}[]

\newtheorem*{Theorem*}{Theorem}

\makeatletter
\renewcommand{\thefigure}{\@arabic\c@figure}
\makeatother
\usepackage{xr}
\usepackage{wrapfig}

\begin{document}

\title{Quantifying assays:  A Modeling tale of variability in cancer therapeutics assessed on cancer cells}
\author{Roumen  Anguelov$^{1,4,}$\footnote{Corresponding author} , G Manjunath$^1$, Avulundiah E Phiri$^1$\\
Trevor T Nyakudya$^3$, Priyesh Bipath$^3$, June C Serem$^2$ Yvette N Hlophe$^3$ \\[6pt]
$^1$Department of Mathematics and Applied Mathematics,\\
$^2$ Department of Anatomy, $^3$Department of Physiology,\\
University of Pretoria\\
\{roumen.anguelov, manjunath.gandhi\}@up.ac.za, edwin@aims.ac.za\\
\{trevor.nyakudya, priyesh.bipath, june.serem, yvette.hlophe\}@up.ac.za\\[6pt]
$^4$Institute~of~Mathematics~\&~Informatics,
Bulgarian~Academy~of~Sciences}
\date{}
\maketitle

\begin{abstract}
Inhibiting a signalling pathway concerns
controlling the cellular processes of a cancer cell's viability, cell division, and death. Assay protocols created to see if the molecular structures of the drugs being tested have the desired inhibition qualities often show great variability across experiments, and it is imperative to diminish the effects of such variability while inferences are drawn. In this paper we propose the study of experimental data through the lenses of a mathematical model depicting the inhibition mechanism and the activation-inhibition dynamics.
 The method is exemplified through assay data obtained from the study of inhibition of the CXCL12/CXCR4 activation axis for the melanoma cells.
To mitigate the effects of the variability of the data on the cell viability measurement, the cell viability is theoretically constructed as a function of time depending on several parameters. The values of these parameters are estimated by using the experimental data. Deriving approximation for the cell viability in a theoretically pre-determined form has the advantages of (i) being less sensitive to data variability (ii) the estimated values of the parameters are interpreted directly in the biological processes, (iii) the amount of variability explained via the approximation validates the quality of the model, (iv) with the data integrated into the model one can derive a more complete view over the whole process. These advantages are demonstrated in the step-by-step implementation of the outlined approach.
 \end{abstract}



\section{Introduction}

To increase the fraction of published ``discoveries" that can be replicated in future investigations
it is desirable to raise the robustness in drawing inference from these experiments.
This is of particularly great importance in drug discovery where the robustness is related to the measured pharmacogenomic response to it.
Here, we develop methods to diminish the potential adverse effects of the inevitable variability in the data from such  \textit{in vitro} experiments through mathematical modeling of certain underlying biological processes.

The value of mathematical models that incorporate biological mechanisms in their formalism for assisting  researchers to plan experiments and shedding light on the underlying mechanism of disease progression is already widely accepted. There is a current trend to combine laboratory-based-research and computational design based on mathematics modeling (e.g., \cite{vera2021mathematical}). For example, differential equation models describe growth rates  in different cellular environments (for e.g., \cite{mendoza2012mathematical,benzekry2014classical, lima2016selection}). It is also common in such studies to fit parameters in the model using the available experimental data to validate the model.  Further, based on mass transport processes involved in a drug release, a strategy is usually made to determine the dosage of the drug required  (e.g., \cite{hixson1931dependence, korsmeyer1983mechanisms, peppas2014mathematical}).

In contrast to understanding the influence of a drug on the macro-scale growth like that of a tumour, in this article, we propose a mathematical model of the underlying biological processes in the environment of a cancer cell.  The aim is to extract the information from the assay data into the mathematical model so that the cell-viability remains less sensitive to the variability in the data. This integration of data with existing biological knowledge embedded in the model provides for, on the one hand, reliably establishing trends of interest and, on the other hand, a better understanding of the biological processes by more detailed interpretation of the data, e.g. identifying the main drivers of the observed dynamics.

During the initial stages of drug discovery,  a cytotoxicity assay such as the crystal violet (e.g. \cite{CVA2016}) assay, as an investigative procedure to study the mechanism of cancer cell inhibition, is an important initial step. In cancer drug discovery, methods are created to see if the drugs being tested contain the desired qualities. In particular, these drugs or compounds are tested for their inhibitory efficacy to cancer cell growth.

At the intuitive level, we can
expect a certain level of variability of the assay data as an inherent property of the conducted experiments. An essential characteristics of assay methods is that the measurement of a cell population results in its destruction. In the crystal violet assay \cite{CVA2016}, cells are fixed in their state, treated with dye, the dye is solubilized and the absorbency of the resulting solution is measured. The fact that no two measurements of the same population can ever be made is a possible reason for the variability in the obtained measurements.
 Hence, an experimental series is never anticipated to be an exact replication of another one.

Statistical-based study of measurements in cell inhibitory assays
to show how the drug responses are different for the same drug across assays has been done previously in  \cite{haibe2013inconsistency}.
There have been counter attempts to draw inferences of consistency from the measurements (e.g.\cite{geeleher2016consistency}), the consistency/inconsistency conclusions have been based on a different statistical reasoning. We note that our approach here is not to dispute or defend the statistical inconsistency, but rather deal with the biological phenomena that could cause variability while inferences are drawn from assay data.

Theoretically, we can formulate the general setting of the experiments in the following abstract form. The population of real interest, e.g., cancer cells in a patient, is not accessible for the experiments. The measurements are in fact from different populations of the same type of cancer cells, with only one measurement per population possible. These populations are similar to each other and to the population of interest, but not exactly the same.

The focus of this work is on the inhibition of cell viability of melanoma cells by blocking the reaction pathway activated by the CXCL12 molecule docking on the CXCR4 sensor on the cell membrane, \cite{orimo2005stromal,cardones2003cxcr4,wong2008translating}. We follow the following protocol of analysis:
\begin{itemize}
\item[(1)] Construct a mathematical model representing the inhibition mechanism and the activation-inhibition dynamics resulting from introducing an inhibitor into the system.
\item[(2)] Construct theoretically the variable which is measured, namely the cell viability to explain the variability in experimental data. For the considered case of inhibition of melanoma, the cell viability is a function depending on four parameters and time.
\item[(3)] Estimate the values of the parameters via least squares fitting of the theoretical viability function to the experimental data.
\item[(4)] Interpret the obtained approximation of the parameters and the respective function in the setting of the model.
\item[(5)] Use the approximation to obtain further information of interest, e.g. $IC_{50}$ curve of concentration of the inhibitor versus time.
\end{itemize}

Of course, a benefit of this additional effort is the increased confidence in the drug-response inferences -- for example the  $IC_{50}$ (or a $IC_{30}$ or a $IC_{70}$) curves  are determined by modelling of biological processes and using the entire data set for the identification of the values of the involved parameters. In contrast, different statistical regression mechanisms can produce a wide range of $IC_{50}$ values especially when there is  great variability in data (e.g., \cite{khinkis2003optimal}).

The remainder of this paper is organized as follows. In Section~\ref{Sec_ID}, we introduce a mathematical model to capture the inhibition dynamics and the resultant effect on the cell viability. The model is derived as a set of ordinary differential equations and analysed as a dynamical system. To facilitate a first reading of the paper, the technical proof of the main theorem is moved to the Appendix. The conducted experiments under the crystal violet assay protocol using L-Kynurenine as an inhibiting agent are described in Section~\ref{Sec_AM}. The graphical representation of the data demonstrates both the variability of data and the existence of a trend. Using the mathematical  model we construct in Section~\ref{Section_4} the theoretical cell viability variable in a specific form of a function of time depending on several parameters. Section~\ref{Section_5} deals with fitting the theoretical cell viability to the set of data as well as some interpretation of the obtained results. Application to deriving $IC_{50}$ for all times within the time range of the experiments is presented in Section~\ref{Section_6}. It is further shown that the method can produce the inhibiting concentration as functions of time for any other level of cell viability, e.g. 30\% or 70\%. Concluding remarks and some questions are discussed in the last section.

\section{Activation-inhibition dynamics} \label{Sec_ID}

\begin{wrapfigure}{r}{0.4\textwidth}
\centering
\includegraphics[width=0.35\textwidth]{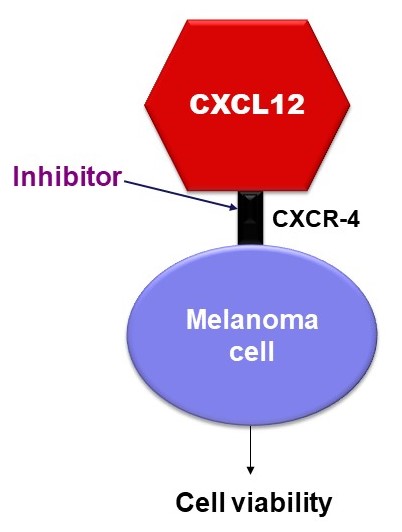}
\caption{Blocking the CXCL12/CXCR4 axis}
\label{inhibit}
 \end{wrapfigure}

Since more cancer cells are expected to die in the presence of an inhibiting drug, the amount of inhibition in the micro-environment of a cancer cell would also be expected to vary with time. We introduce a mechanistic model that deals with the inhabitancy of the inhibiting molecules on the cancer cell.
Melanoma cells express chemokine receptor 4 (CXCR-4) located on the cell membrane \cite{wong2008translating}. When the chemokine ligand 12 (CXCL12) bounds to CXCR4 it activates signalling pathways such as mitogen-activated protein kinase (MAPK) and phosphatidylinositol 3-kinase (PI3K). The activated signaling pathways promote cancer cell proliferation, migration, and adhesion. Hence, the CXCL12/CXCR-4 axis is crucial in cancer metastasis.

The investigation, both qualitatively and quantitatively, of the temporal dynamics of blocking this axis, is the focus of this section, see Figure \ref{inhibit}.

\subsection{Mathematical model}

Depending on which agent, the activator CXCL12 or an inhibitor, docks on the sensor CXCR4 at each time instant, the signaling pathway for producing essential for the survival and proliferation of the cell is either activated or not. The dynamic interaction between activation and inhibition, an essential factor determining the viability of the population, is captured through a system of differential equations. For simplicity, we use terms that are not technical so that the model can be understood by researchers in different disciplines. We  introduce the variables involved in our model:

$L$ - total number of free (not docked) activating molecules.

$X$ - total number of free (not docked) molecules of the inhibiting substance.

$R$ - total number of unoccupied ``docking" places on the sensors where $X$ or $L$ can attach

$P$ - total number of docking places on the sensors occupied by activating molecules $L$.

$Q$ - number of docking places on the sensors occupied by the inhibiting substance molecules $X$.

\begin{figure}[t]
\centering
\includegraphics[width=16cm]{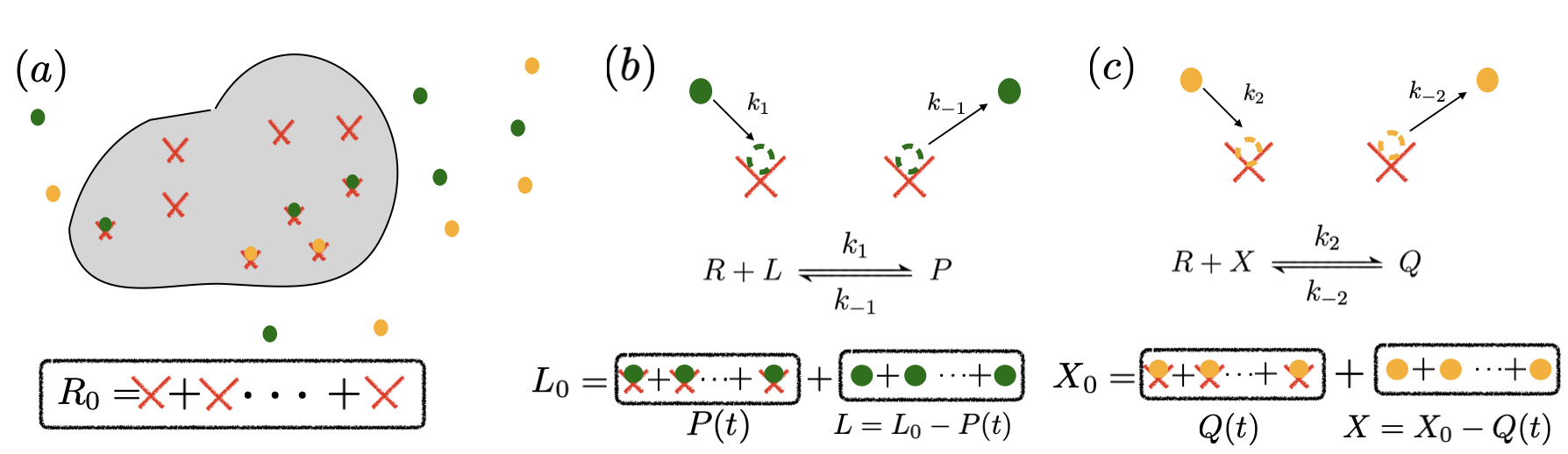}
\caption{Kinetics: docking places (Red), activating molecules (green) and inhibiting molecules (yellow) (a). Sensors (docking places) on a cancer cell (b) \& (c) Chemical reactions along with the relationship between dynamic and static quantities.}
\label{Fig_MK}
 \end{figure}

All these quantities are dynamic in the sense they vary with time.
A summary of how these quantities interact is shown in Fig.~\ref{Fig_MK}.

The activation and blocking of the sensors  can be described by  equations similar to that used in chemical reactions:
\begin{eqnarray}
&&R+L\  \autorightleftharpoons{\hspace{2mm}$k_1$\hspace{2mm}}{$k_{-1}$} \ P \label{RL}\\
&&R+X\ \autorightleftharpoons{\hspace{2mm}$k_2$\hspace{2mm}}{$k_{-2}$} \ Q\label{RX}
\end{eqnarray}
The constants $k_1$ and $k_{-1}$ are specific for the activator, namely the considered here CXCL12 molecule, while $k_2$ and $k_{-2}$  are specific for the inhibiting agent.  These reactions can be represented by the following differential equations using the principle of mass action reaction kinetics:
\begin{eqnarray}
\frac{dP}{dt}&=&k_1RL-k_{-1}P\label{eqP}\\
\frac{dQ}{dt}&=&k_2RX-k_{-2}Q,\label{eqQ}\\
\frac{dL}{dt}&=&-k_1RL+k_{-1}P,\label{eqL}\\
\frac{dX}{dt}&=&-k_2RX+k_{-2}Q,\label{eqX}\\
\frac{dR}{dt}&=&-k_1RL-k_2RL+k_{-1}P+k_{-2}Q.
\end{eqnarray}
By adding the appropriate equations we obtain
\begin{equation}\label{eqConstants}
\frac{d(P+L)}{dt}=\frac{d(Q+X)}{dt}=\frac{d(P+Q+R)}{dt}=0.
\end{equation}
Hence, the quantities $P+L$, $Q+X$, $R+P+Q$ remain constant during the reactions (\ref{RL})-(\ref{RX}). This should not be surprising since these reactions do not consider any growth or decay in the total number of molecules of the activating agent, the inhibiting substance and docking places on the sensors. We have
\begin{eqnarray*}
L_0&:=&P+L\ = \ \text{total number of activating molecules,  docked or free;}\label{eqL_0}\\
X_0&:=&Q+X\ =\ \text{total number of molecules of the inhibiting substance, docked or free;\hspace{1cm}}\label{eqX_0}\\
R_0&:=&P+Q+R\ =\ \text{total number of docking places on the sensors occupied by}\\
&&\hspace{2.5cm} \text{activating molecules, occupied by the inhibiting substance or free;}\label{eqR_0}
\end{eqnarray*}

It is possible to consider the dynamics of generation and destruction of the activating molecules as well as destruction and re-supply of the inhibiting substance. However, considering these quantities as constants may be sufficient as a first step as well as relevant to the conducted experiments due to their limited time span.

Substituting $L=L_0-P$, $X=X_0-Q$ and $R=R_0-P-Q$ in to equations (\ref{eqP})-(\ref{eqQ}) we obtain system of two equations only:
\begin{eqnarray}
\frac{dP}{dt}&=&k_1(R_0-P-Q)(L_0-P)-k_{-1}P,\label{eqP2}\\
\frac{dQ}{dt}&=&k_2(R_0-P-Q)(X_0-Q)-k_{-2}Q.\label{eqQ2}
\end{eqnarray}

\subsection{Equilibrium of the inhibition dynamics}

We describe the qualitative dynamics of the system of differential equations \eqref{eqP2}--\eqref{eqQ2}. The only set of initial conditions that are physically plausible is the set
\begin{equation} \label{eq_D}
D : = \{(P,Q): 0 \le P \le L_0 \: \: \& \: \: 0 \le Q \le X_0 \}.
\end{equation}

The following theorem shows that for all solutions of \eqref{eqP2}--\eqref{eqQ2}initiated in $D$, the activation level, as given by the value of $P$ and the inhibition level, as given by the value of $Q$, approach an equilibrium as time increases.  The practical message is that if the assumed conditions remain prevalent in the long term, the inhibition level approaches a constant.

\begin{Theorem} \label{Thm_eqm}
Consider the dynamical system \eqref{eqP2}--\eqref{eqQ2} with all parameters being positive as well as the set $D$ as defined in \eqref{eq_D}. Then the following results hold:\vspace{-3mm}

\renewcommand{\labelenumi}{(\roman{enumi})}
\begin{enumerate}
\item There is a unique equilibrium point $(P^*,Q^*)$ in the interior of $D$.

\item The equilibrium $(P^*,Q^*)$ is a stable proper node and hence locally asymptotically stable.

\item The set $D$ is positively invariant, that is for any initial condition $(P(0),Q(0))\in D$ the solution $(P(t),Q(t))$ of the system \eqref{eqP2}--\eqref{eqQ2} is defined for all $t>0$ and $(P(t),Q(t))\in D$, $t>0$.

\item The set  $D$ is contained in the basin of attraction of $(P^*,Q^*)$.

\end{enumerate}\vspace{-3mm}
\end{Theorem}

\begin{figure}[ht]
\centering
\includegraphics[trim=18.5cm 0cm 19cm 0cm,clip, width=9cm]{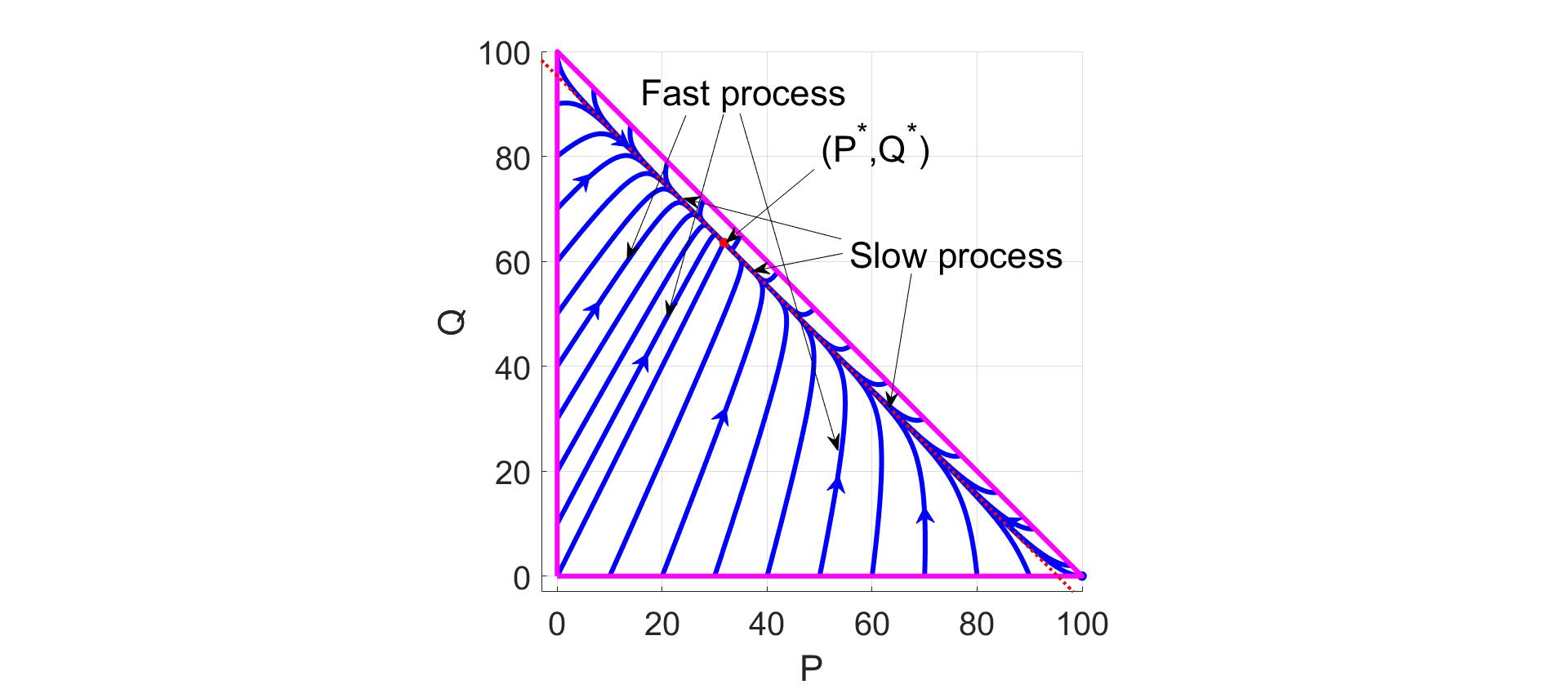}
\caption{The domain $D$ is the set with the purple boundary, and the phase diagram with $(R_0,L_0,X_0,k_1,k_2,k_{-1},k_{-2}) = (100,100,200,10,10,1,1)$}
\label{Fig_phaseLin}
 \end{figure}

Note that the statement (iv) in Theorem~\ref{Thm_eqm} is stronger than saying $D$ is positively invariant, while statement (ii) or (iii) does not specify the basin of attraction of $(P^*,Q^*)$. The proof of Theorem~\ref{Thm_eqm} is presented in the Appendix.

The phase diagram on Figure \ref{Fig_phaseLin} shows how the trajectories of the solutions of the system (\ref{eqP2})-(\ref{eqQ2}) approach the equilibrium $(P^*,Q^*)$. One can observe that, as stated in Theorem \ref{Thm_eqm}, all trajectories converge to $(P^*,Q^*)$. One can further observe that all trajectories initiated at the boundary of $D$ excluding the origin are tangential to the line
\begin{equation}\label{lineSlow}
P+Q=P^*+Q^*
\end{equation}
at the equilibrium. More precisely, the trajectories first approach the mentioned line (\ref{lineSlow} and then continue towards the equilibrium while getting closer to the line and becoming visually indistinguishable from it. This type of dynamics occurs when the Jacobian of the vector field at the equilibrium has two distinct negative eigenvalues as captured in the concept \textit{stable proper node} \cite[Section 1.4]{Tabor}. The next subsection discusses the physiological and the mathematical aspects of such dynamics.

\subsection{Fast and slow manifolds}

We can rewrite the system(\ref{eqP2})-(\ref{eqQ2}) in the form
\begin{eqnarray}
\frac{dP}{dt}&=&k_1R_0\frac{R_0-P-Q}{R_0}(L_0-P)-k_{-1},P\label{eqP3}\\
\frac{dQ}{dt}&=&k_2R_0\frac{R_0-P-Q}{R_0}(X_0-Q)-k_{-2}Q.\label{eqQ3}
\end{eqnarray}
We next set $\tilde{k}_{i}=k_{i}R_0$, with $i=1$ or $i=2$ and along with $k_{-1}$,$k_{-2}$ these quantities are all rates where the measuring unit is ``per unit time". Specifically,  $\tilde{k}_{1}$ is the rate of attachment of the activating agent and $\tilde{k}_{2}$ is the rate of attachment of the inhibiting agent when all docking places are free ($P=Q=0$). Typically in the described type of physiological processes and as it is the case with activation (by CXCL12) or blocking (by an inhibiting agent, e.g. L-kynurenine) of the receptor CXCR4, we have
\begin{equation}\label{waygreater}
\tilde{k}_1\gg k_{-1}, \ \ \ \tilde{k}_{2}\gg k_{-2}.
\end{equation}
The attachment rate is reduced by the fraction of available docking places, that is
$\displaystyle\frac{R_0-P-Q}{R_0}$ (dimensionless). When many docking spaces are available, that is $P+Q$ is small compared to $R_0$, we have a fast process of attachment (growth of both $P$ and $Q$) driven by $\tilde{k}_{1,2}$. When the docking places are near full, the rates of change in $P$ and $Q$ depend on the rate new docking spaces become available, that is $k_{-1}$ and $k_{-2}$. Considering (\ref{waygreater}), this is relatively much slower process.

The fast and the slow processes are represented mathematically by the invariant manifolds associated with the eigenvalue of the Jacobian of the right-hand side of the system \eqref{eqP2}--\eqref{eqQ2}. As shown in the proof of Theorem \ref{Thm_eqm} (see Appendix), the Jacobian has two distinct negative eigenvalues. Let us denote them by $-\lambda$ and $-\mu$, where $\mu>\lambda>0$. It follows easily from the invariant manifold theory, e.g. \cite[Theorem 3.2.1]{Wiggins}, that there exists locally a one- dimensional invariant manifold corresponding to each eigenvalue. More specifically each invariant manifold is a smooth curve passing through the equilibrium and tangent to the eigenvector of the respective eigenvalue. The invariantness  further implies each of these curves is a trajectory of a solution. Any non-equilibrium solution as time decreases can be extended to intersect the boundary of $D$ in view of Theorem~\ref{Thm_eqm}. The manifold (curve) corresponding to the eigenvalue $-\lambda$ represents the fast process, while the manifold (curve) corresponding to the eigenvalue $-\mu$ represents the slow process. Hence, we refer to them as the fast manifold and the slow manifold, respectively. Any solution which is not initiated on any of the two manifolds displays features of both, with the dynamics on the slow manifold eventually dominating. Hence, the trajectories of all solutions except for those on the fast manifold are tangential to the slow manifold at the equilibrium.

In the special case when
\begin{equation}\label{ratesequal}
k_1=k_2,\ \ \  k_{-1}=k_{-2}
\end{equation}
and using the results in the Appendix, one can derive the values of $\lambda$ and $\mu$ in the following explicit form
\begin{equation}\label{eigenvalues}
\lambda=k_1(R_0-P^*-Q^*), \ \ \mu=\lambda +k_1(L_0+X_0-P^*-Q^*).
\end{equation}
Clearly, $\lambda<\mu$. Further, it is easy to show that
\begin{eqnarray}
\lambda&<&k_{-1}\left(1+\frac{R_0}{L_0+X_0-R_0}\right),\label{boundLambda} \\
\mu&>&\lambda+\tilde{k}_1\frac{L_0+X_0-R_0}{R_0}.\label{boundMu}
\end{eqnarray}
The upper bound of $\lambda$ in \eqref{boundLambda} is directly proportional to $k_{-1}$ indicating that $\lambda$ is small when $k_{-1}$ is small. Similarly the lower bound of $\mu$ in \eqref{boundMu} indicates that $\mu$ is significantly larger than $\lambda$ when $\tilde{k}_1$ is large. One should note that increasing $X_0$ does not change the situation. In fact, the upper bound of $\lambda$ decreases with respect to $X_0$, while the lower bound of $\mu$ increases, making the difference between the fast and the slow processes even more pronounced.

\begin{figure}[ht]
\centering
\includegraphics[width=9cm]{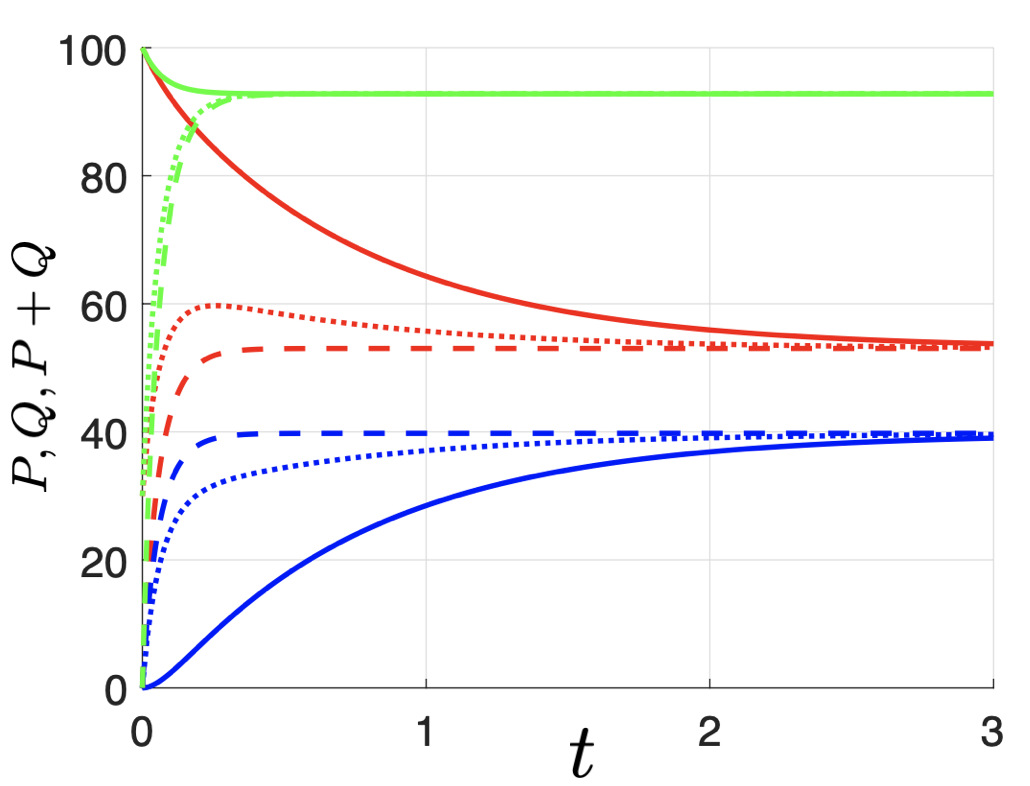}
\caption{Graphs of $P$ (red), $Q$ (blue) and $P+Q$ (green) with  $(R_0,L_0,X_0,k_1,k_2,k_{-1},k_{-2}) = (100,100,200,10,10,1,1)$ corresponding to different set of initial conditions.}
\label{Fig_Conv}
 \end{figure}

In this specific case, both the slow manifold (associated with $\lambda$) and the fast manifold (associated with $\mu$) are straight lines with respective equations
\begin{eqnarray}
&&P+Q=P^*+Q^*,\label{eqslowline}\\
&&Q^*P-P^*Q=0.
\end{eqnarray}
These straight lines are visible on Figure \ref{Fig_phaseLin}, where the parameters are such that \eqref{ratesequal} holds. The temporal dynamics are illustrated on Figure \eqref{Fig_Conv}. Compared to the graphs of $P(t)$ and $Q(t)$, the graph of $P(t)+Q(t)$ converges to its equilibrium much faster. Considering \eqref{eqslowline}, this indicates the convergence of the solutions $P(t)$, $Q(t)$ to the slow manifold which is much faster than the convergence to   their equilibrium values. This is a different way of representing the property that \vspace{-5mm}

\begin{equation}\label{prop}
\begin{aligned}&\textit{\hspace{-3mm}all solutions, excluding the one initiated at the origin, eventually approach the}\\[-4pt] &\textit{\hspace{-3mm}equilibrium of the system on a trajectory indistinguishable from the slow manifold.}
\end{aligned}
\end{equation}\vspace{-5mm}

\begin{figure}[ht]
\centering
\includegraphics[trim=18.5cm 0cm 19cm 0cm,clip, width=9cm]{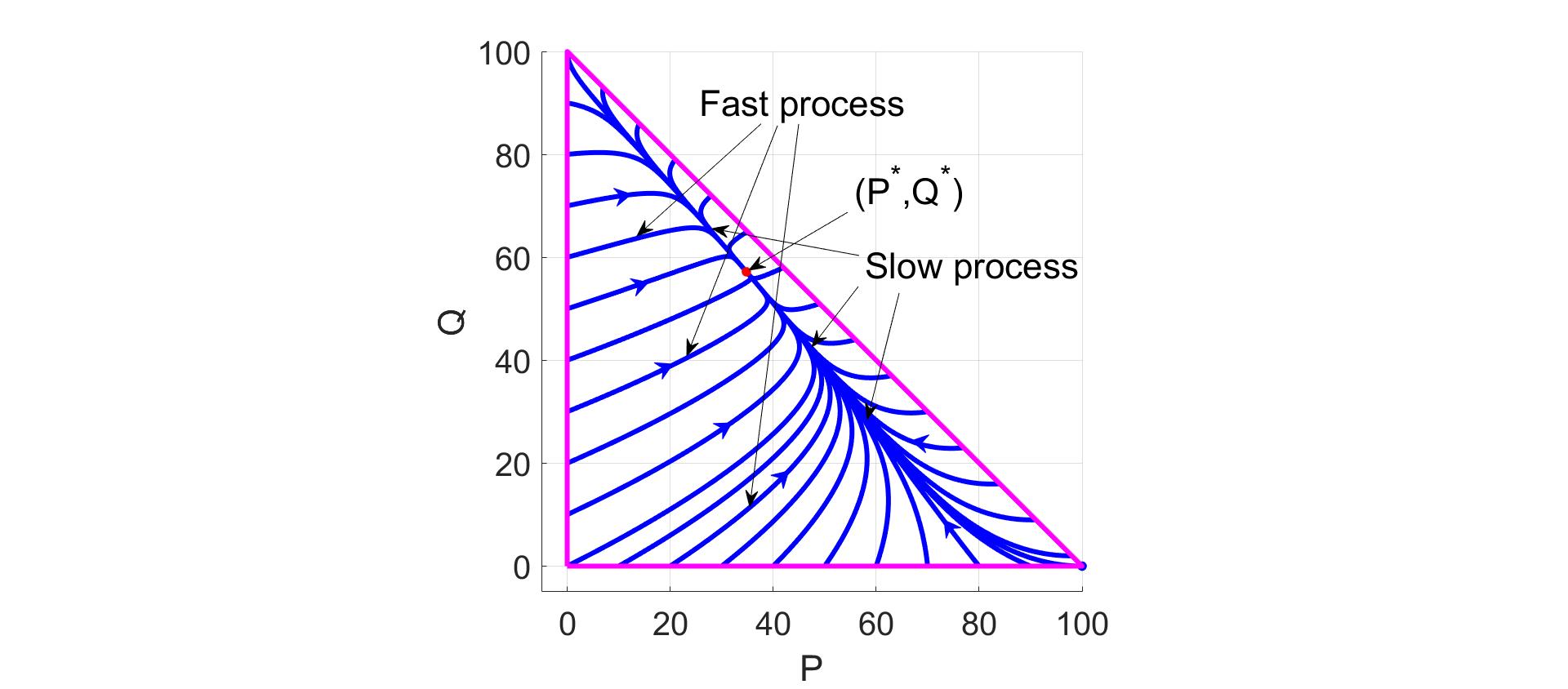}
\caption{The domain $D$ is the set with the purple boundary, and the phase diagram with $(R_0,L_0,X_0,k_1,k_2,k_{-1},k_{-2}) = (100,100,200,20,5,3,1)$}
\label{Fig_phaseNonlin}
 \end{figure}

The special case when \eqref{ratesequal} was considered in some detail to illustrate the ideas that extend to the general case, i.e.,  when \eqref{ratesequal} is not true. The behavior of the solution in the general case is determined (as stated in \eqref{prop}) by a fast and slow manifold associated with the equilibrium as long as \eqref{waygreater} holds. Figure \eqref{Fig_phaseNonlin} represents a phase diagram for different values of the rate constants where \eqref{ratesequal} does not hold. The fast manifold, the slow manifold, and the property \eqref{prop} can  be observed in the figure.

\section{Assay measurement of the viability of melanoma cells} \label{Sec_AM}

The inhibitor compound used in this study is L-Kynurenine (L-Kyn). L-Kyn, a downstream metabolite of the amino acid tryptophan, has been shown to inhibit proliferation and induce cell death of melanoma cell-lines via \textit{in vitro} studies \cite{walczak2020effect}. In addition to being an endogenous derivative, L-Kyn is thereby a biologically suitable compound to be tested via melanoma-related \textit{in vitro} assays and to demonstrate a reduction in cell viability at increasing concentrations while still being tolerated by non-cancerous cells \cite{marszalek2021kynurenine}.

Melanoma cells were exposed to L-Kyn at 1-4 milli molars (mM) for 24,48 and 72 hours. The cells were then tested for cell viability using the crystal violet assay. The goal is to measure the effect of the inhibitory agent on cell viability. Theoretically, the cell viability of a treated population is a function of time. At any given time it is defined as the ratio of the treated population size over the size of this population if untreated.  Following the crystal violet assay, \cite{CVA2016}, the cell viability is measured as follows. A set of wells  is prepared with 5000 cell each in growth cell culture medium and incubated at physiological conditions. The size of the population is verified via manual counting. After one day all cells are attached to the wells. The wells are divided into groups: control (no inhibition), positive control (NOC), and the other groups treated with  varied concentrations of the inhibiting agent. At least 9 wells from each group are analyzed in 24 hours, 48 hours, and 72 hours. The cells  are fixed in their state, treated with crystal violet dye (absorbed only by the DNA of living cells), and the wells are washed so that only the attached cells remain. Then the absorbance of the dye is measured. The cell viability is calculated as the ratio of the absorbance of the treated population over the absorbance the control population.

\begin{figure}
\centering
\begin{minipage}{0.5\textwidth}
\includegraphics[width=\textwidth]{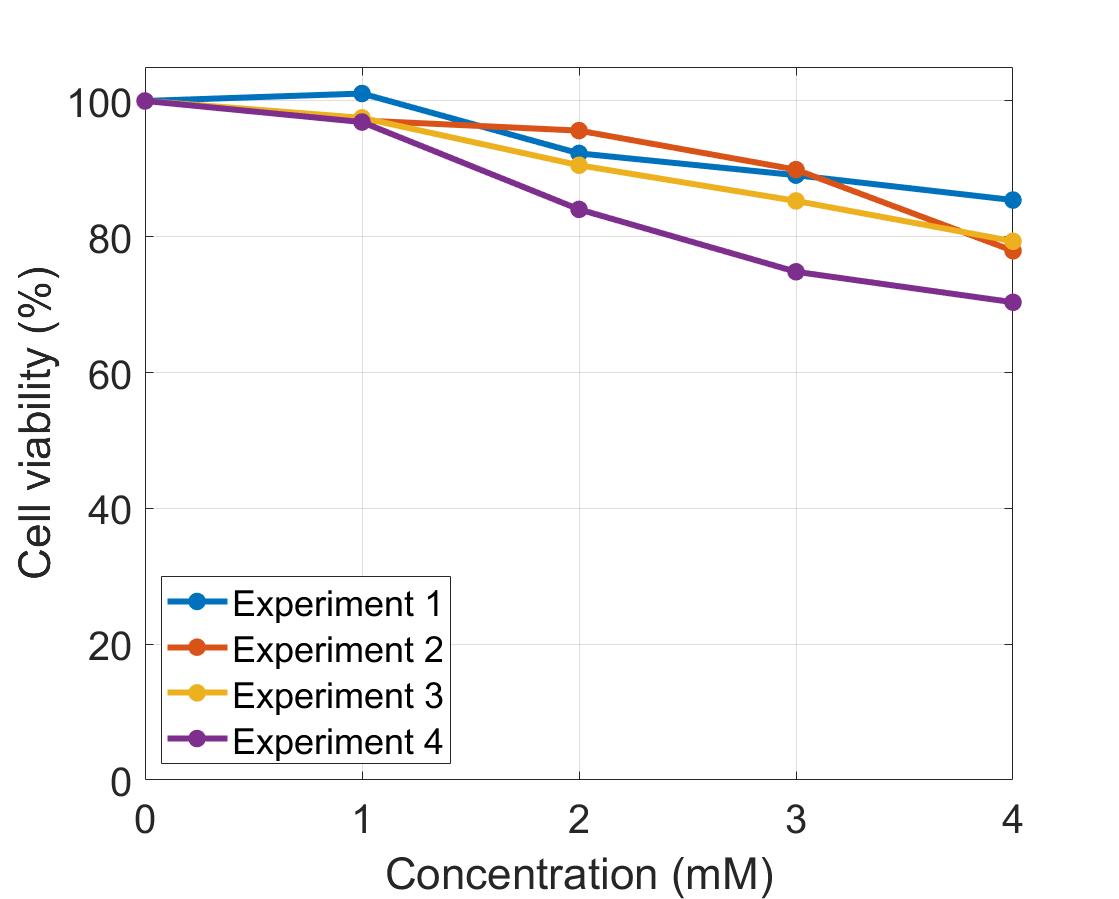}
\caption{Cell viability at 24 hours}
\label{Via24}
\end{minipage}%
\begin{minipage}{0.5\textwidth}
\includegraphics[width=\textwidth]{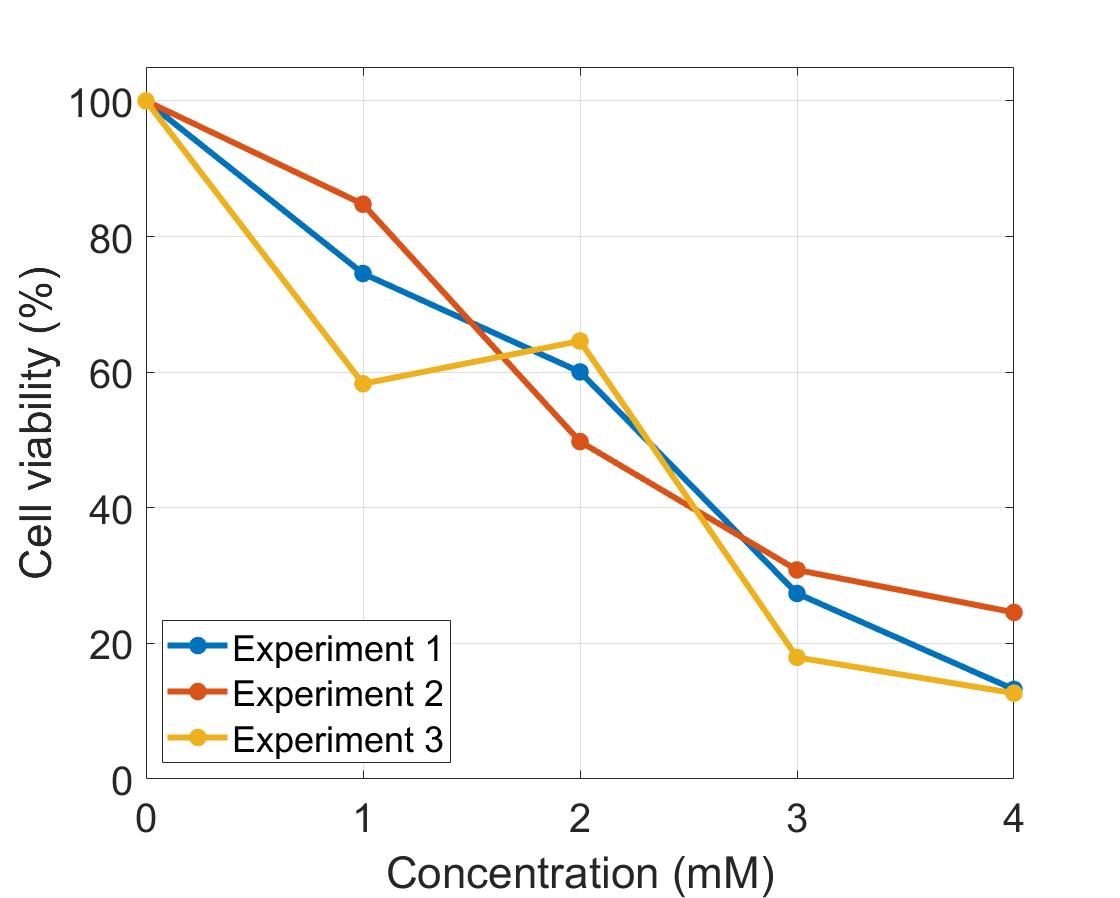}
\caption{Cell viability at 48 hours}
\label{Via48}
\end{minipage}

\begin{minipage}{0.5\textwidth}
\includegraphics[width=\textwidth]{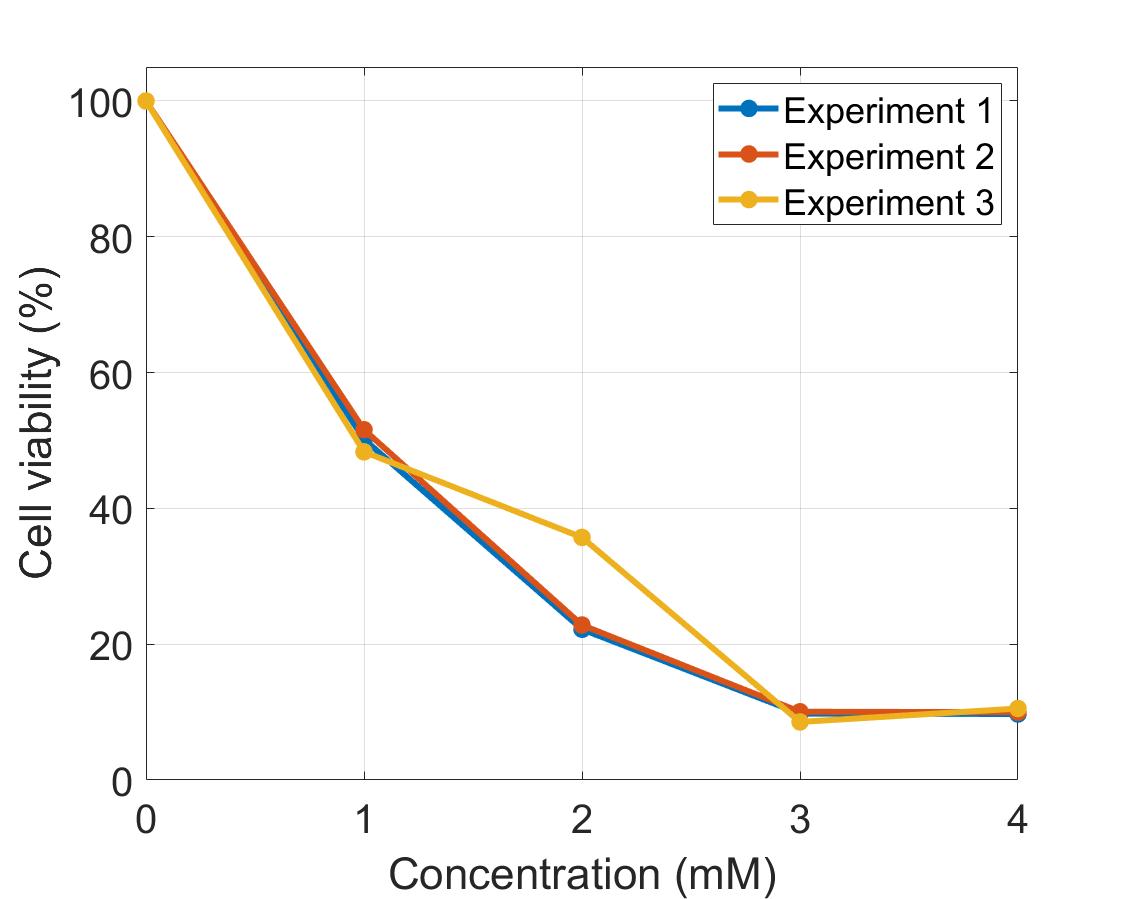}
\caption{Cell viability at 72 hours}
\label{Via72}\vspace{5mm}
\end{minipage}%
\begin{minipage}{0.5\textwidth}
\includegraphics[width=\textwidth]{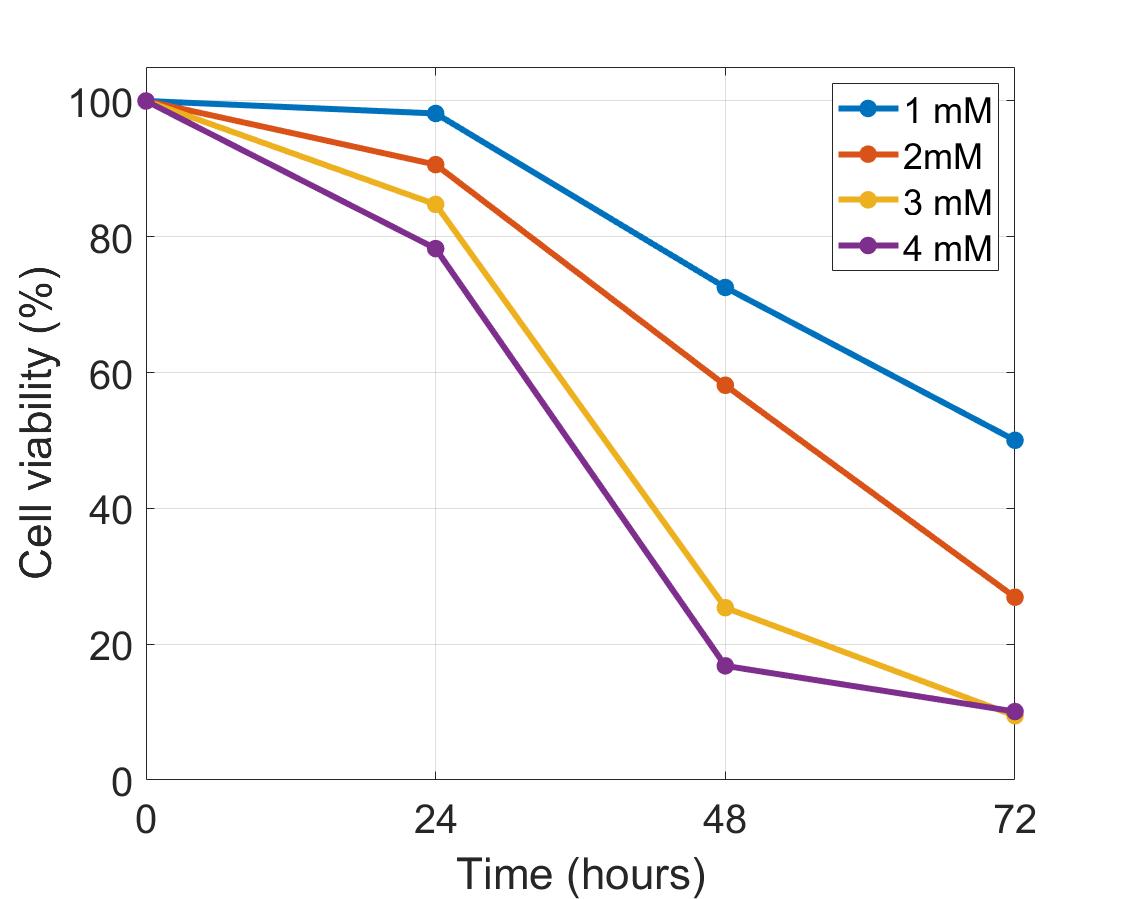}
\caption{Average cell viabilities over time for the different concentrations of L-Kyn}
\label{AveOverTime}
\end{minipage}
 \end{figure}

The main limitation on the accuracy of the measurement is that the measurement of any population results in its destruction. The measurements are from populations that are expected to be similar, yet they are not the same. Variations are observed in each analyzed set of treated populations as well as in the populations in the negative control group, that is untreated. The calculated cell viability versus concentration of L-Kyn is given in Figures \ref{Via24}, \ref{Via48}, \ref{Via72} for the specified measurement times. Every point is calculated using the average of three wells. The percentage is taken with respect to the control populations so that the control population is always 100\%. At least three experiments were conducted. The graphs on Figures \ref{Via24}, \ref{Via48}, \ref{Via72} demonstrate significant variability of the results from one experimental series to the next. The main reason for this variability is that no two measurements can be taken from the same population.  However, trends can be extracted when considering large number of experiments. The simplest way to integrate data at the same time under the same treatment is by using the mean. Figure \ref{AveOverTime} represented the average cell viability versus time for the three concentrations in the experiments. One can observe that there are general trends of decreasing of cell viability with respect to time and with respect to concentration. Quantifying these general trends only from the data, particularly considering the mentioned variability and some apparent exceptions from these trends, is not likely to be very reliable. In the sequel we suggest a method of integrating the data over time using the model \eqref{eqP2}-\eqref{eqQ2}. In this way, the data is considered in its entirety over time and not only at individual time instances.

\section{Dynamics of the cell viability} \label{Section_4}

We associate the data discussed in Section 3 with the model \eqref{eqP2}-\eqref{eqQ2} in the following way. In this section we construct mathematically the concept of cell viability as a function of time using the theoretical definition discussed already. In the sections that follows, we consider the assay data as approximate measurements of the value of this function at given times, so that reliable and accurate approximation of this function is derived.

Let $M(t)$ denote the size of a natural population of cells.  Suppose that sufficient resources and optimal environment (as in the assay) are provided. Under such condition one may assume a constant growth rate model for the population, that is $M(t)$ is a solution of the differential equation
\begin{equation}\label{eqM}
\frac{dM}{dt} = rM,
\end{equation}
where $r>0$ is the constant (relative) growth rate. Then we have
$$M(t) = M(0) e^{rt}.$$

Let $\widehat{M}(t)$ denote the size of an identical at $t=0$ population of cells, which is subjected to some treatment. In the setting of the model \eqref{eqP2}-\eqref{eqQ2}, the treatment is inhibition through the blocking of the CXCR4 sensor. Then the cell viability is the function
\begin{equation}\label{TheoreticalCellViability}
\frac{\widehat{M}(t)}{M(t)},
\end{equation}
where the fraction is expressed as a percentage. For every considered concentration, the data in Section 3 can be considered as a set of measurements of this function at times $t=0$, $t=24$, $t=48$, $t=72$ hours. We can interpret the error of the measurements in different ways.  In whichever way we consider the measurements and the error associated with them, the goal is to derive an approximation to the function \eqref{TheoreticalCellViability} which agrees best with the data. For that goal, first we derive from the model \eqref{eqP2}-\eqref{eqQ2} a suitable representation of \eqref{TheoreticalCellViability}.

The impact of the inhibition can be modelled either as a function of $Q$ or as a function of $P$. We chose here $Q$. The inhibition $Q$ is expected to reduce the growth rate. Thus, we have
\begin{equation}  \label{eq_inhib}
	\frac{d\widehat{M}}{dt} = (r-\alpha Q(t))M,
\end{equation}
where $\alpha$ is a positive constant. If the function $Q$ is known, equation \eqref{eq_inhib} can be solved explicitly and we have
\begin{equation}  \label{eqn_cell_viability}
	\frac{\widehat{M}(t)}{M(t)} = \frac{M_0 e^{\displaystyle rt-\alpha \int_{0}^t Q(\theta) \, d\theta}}{M_0 e^{\displaystyle rt}} = e^{\displaystyle -\alpha \int_{0}^t Q(\theta) \, d\theta}.
\end{equation}
Interestingly, the cell viability function does not depend on the constant  $r$, while it depends through $Q$ on the values of $k_1$, $k_{-1}$,  $k_2$, $k_{-2}$, $L_0$, $R_0$, $X_0$, $P(0)$, $Q(0)$ in
\eqref{eqP2}--\eqref{eqQ2} of which only $X_0$, the concentration of the inhibiting substance, is under the gambit of the experimenter.

In the setting of the experiments, we have $Q(0)=0$. However, as discussed earlier, driven by the fast process the solution is very quickly close to the slow manifold. This motivates the validity of approximating such a solution with a solution on the slow manifold, essentially reducing the dimensionality of the system to one, namely a differential equation about $Q$. As an illustration, let us use again the case when \eqref{ratesequal} holds. Then the slow manifold is the straight-line \eqref{eqslowline} and $Q$ satisfies on this manifold the differential equation
\[
\frac{dQ}{dt}=k_1(R_0-P^*-Q^*)(X_0-Q)-k_{-1}Q
\]
with the solution given by
\[
Q(t)=Q^*-(Q^*-Q(0))e^{-\lambda t},
\]
where $\lambda=k_1(R_0-P^*-Q^*)+k_{-1}$.
Hence, the $Q$-coordinate of a solution of the system \eqref{eqP2}--\eqref{eqQ2} can be approximated by
\begin{equation}\label{approx1}
Q(t)\approx Q^*-(Q^*-\bar{Q})e^{-\lambda t},
\end{equation}
where $(\bar{P},\bar{Q})$ is appropriate point on the slow manifold.
\begin{figure}
\includegraphics[trim=0cm 0cm 2cm 0cm,clip,width=0.5\textwidth]{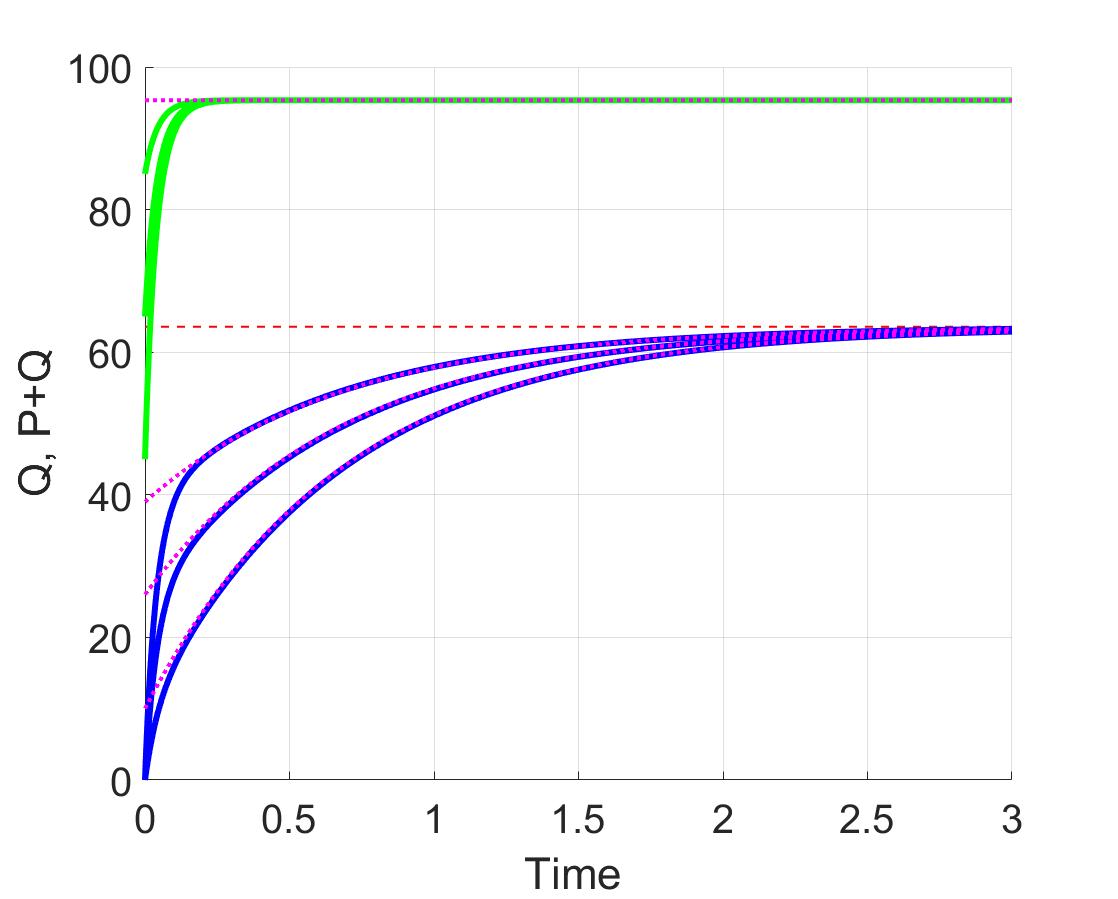}\ \
\includegraphics[trim=0cm 0cm 2cm 0cm,clip,width=0.5\textwidth]{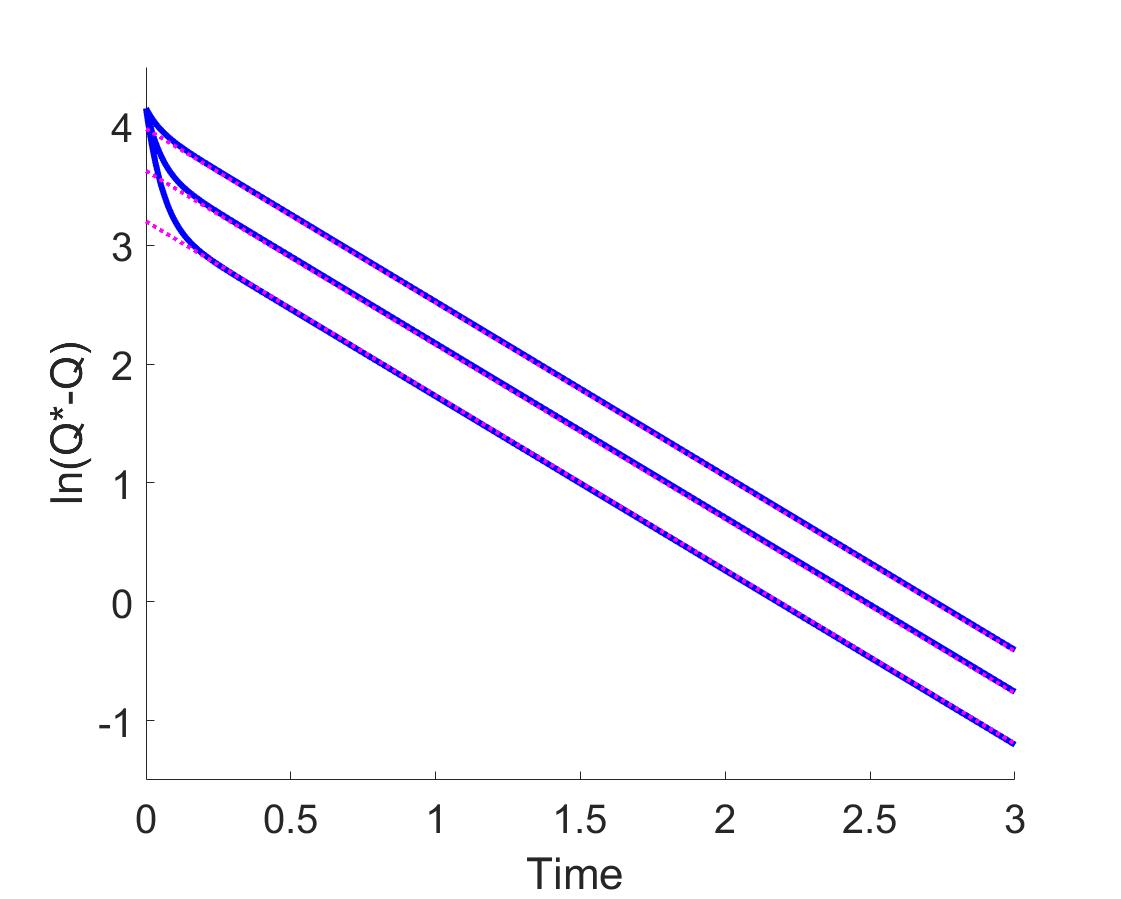}\vspace{-3mm}

(a) \hspace{8cm} (b)

\caption{Solutions from model \eqref{eqP2}-\eqref{eqQ2} with $Q(0)\!=\!0$, $P(0)\in\{45,65,85\}$ and parameters as for Figure \ref{Fig_phaseLin}.\\
(a) Graphs of the $Q$-coordinate (solid blue) of solutions and their approximations on the slow manifold (dotted red). Graphs of $P+Q$ (solid green) for the same solutions.\\
(b) Graphs of $\ln(Q^*-Q)$ for the graphs of $Q$ and their approximation given on (a).}
\label{SlowLineApprox}

\includegraphics[trim=0cm 0cm 2cm 0cm,clip,width=0.5\textwidth]{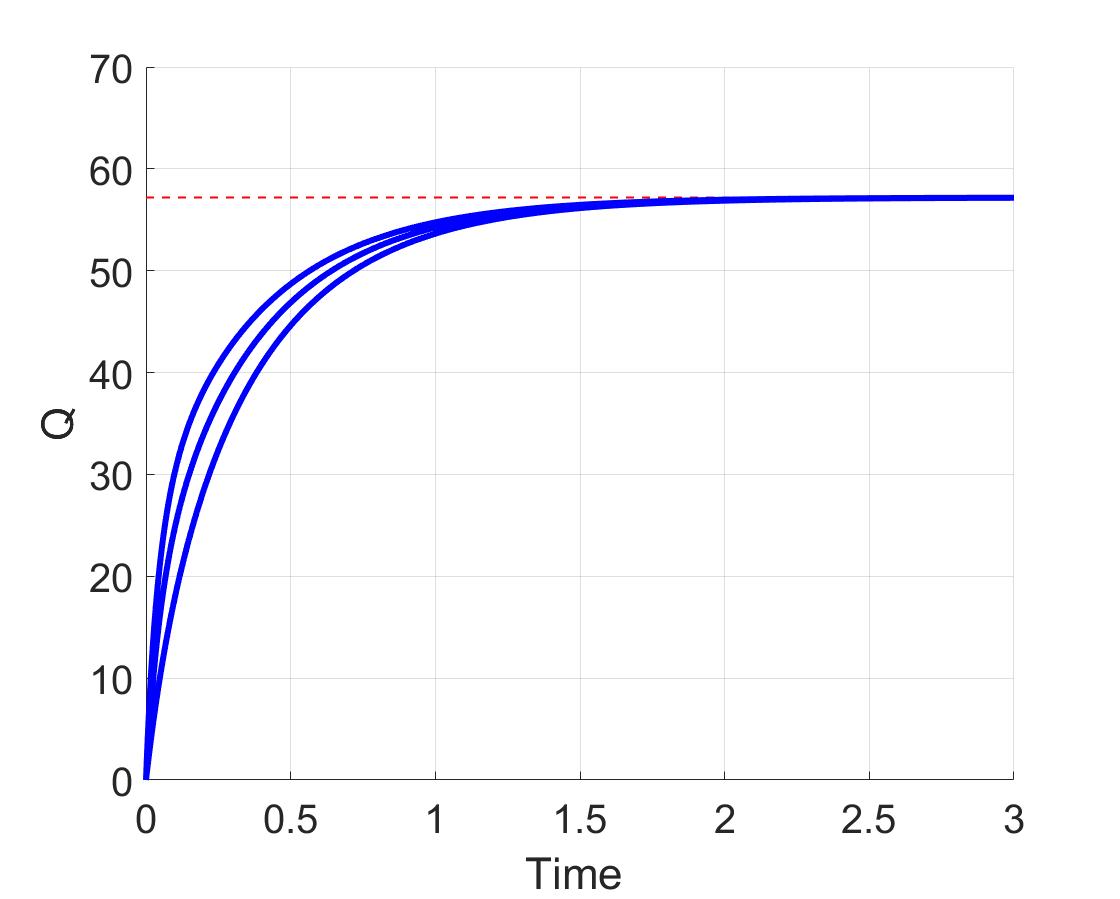}\ \
\includegraphics[trim=0cm 0cm 2cm 0cm,clip,width=0.5\textwidth]{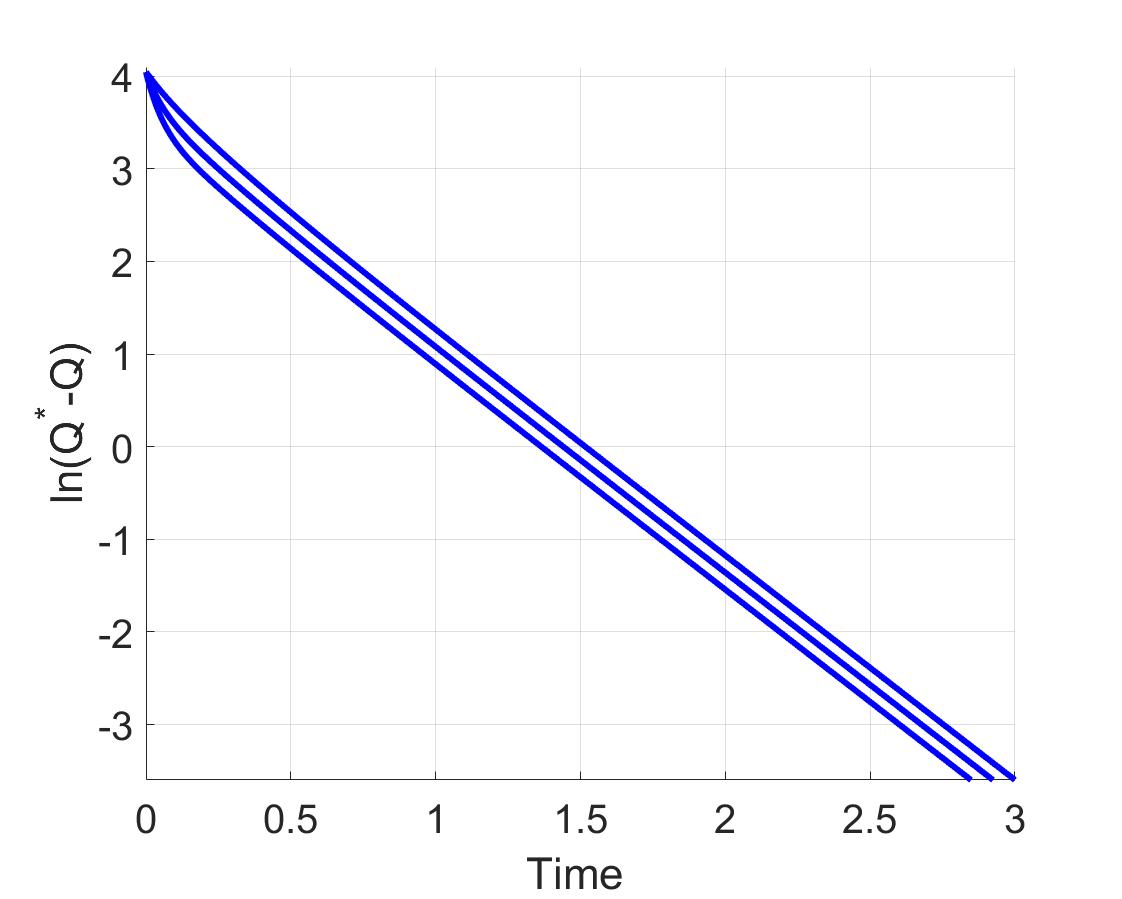}\vspace{-3mm}

(a) \hspace{8cm} (b)

\caption{Solutions from model \eqref{eqP2}-\eqref{eqQ2} with $Q(0)\!=\!0$, $P(0)\in\{25,50,75\}$ and parameters as for Figure \ref{Fig_phaseNonlin}.\\
(a) Graphs of the $Q$-coordinate of solutions.\\
(b) Graphs of $\ln(Q^*-Q)$ for the graphs of $Q$ given on (a).}
\label{SlowNonlinApprox}
\end{figure}

This approximation is illustrated in Figure \ref{SlowLineApprox}(a) for a sample of solutions. One can observe that the graph of the approximation becomes indistinguishable from the $Q(t)$ at about the same time when the graph of $P+Q$ is indistinguishable from its equilibrium, that is the solution is practically on the slow manifold. This happens in a relatively small period of time compared to the time it takes for $Q(t)$ to reach its equilibrium $Q^*$ (the red dashed line). The validity of the approximation of the form \eqref{approx1} can be graphically tested by plotting $\ln(Q^*-Q(t))$. If \eqref{approx1} is valid, then graph of
\[
\ln(Q^*-Q(t))\approx \ln(Q^*-\bar{Q})-\lambda t,
\]
is approximately a straight line. The inverse is also true. If the graph of $\ln(Q^*-Q(t))$ is approximately a straight line, then $Q$ can be approximated as in \eqref{approx1}.

In a general model, where \eqref{ratesequal} does not necessarily hold, approximation of the form \eqref{approx1} in a sufficiently small neighborhood of the equilibrium follows from the Hartman-Grobman Theorem. However, the validity over the whole slow manifold can be tested by plotting the graph of $\ln(Q^*-Q(t))$. Figure \ref{SlowNonlinApprox} (a) represents the graphs of $Q$ for a set of solutions of the model \eqref{eqP2}-\eqref{eqQ2} with parameter values as for Figure \ref{Fig_phaseNonlin}. On Figure \ref{SlowNonlinApprox}(b) the graphs of $\ln(Q^*-Q(t))$ for the same solutions are given. Similar to Figure \ref{SlowLineApprox}(b), these are straight lines except for a small time interval in the beginning. In all our numerical experiments we found this to be true as long as \eqref{waygreater} holds. Hence, we use the approximation \eqref{approx1} with $\bar{Q}$ and $\lambda$ yet unknown.

Then, from \eqref{eqn_cell_viability} we obtain
\begin{equation}  \label{eqn_cell_viability2}
	\frac{\widehat{M}(t)}{M(t)} \approx e^{\displaystyle -\alpha Q^*\frac{e^{-\lambda t}-1+\lambda t}{\lambda}-\alpha\bar{Q}\frac{1-e^{-\lambda t}}{\lambda}}.
\end{equation}
The function in \eqref{eqn_cell_viability2} depends on many unknown parameters. These include $\alpha$ and through $Q^*$ and $\bar{Q}$, all parameters of \eqref{eqP2}-\eqref{eqQ2} as well as $P(0)$. Hence, this function is not intended to be derived from the model, but rather it provides means of assimilating the data as approximate observations of functions of specific form. Taking into account that, as discussed in Section 3, some error is involved in the measurements, we may consider that measurements of $\displaystyle \frac{\hat{M}(0)}{M(0)}$, while close to 1 are not necessarily 1. Then we have for the cell viability an approximation by a function of the form
\begin{equation}  \label{eqn_cell_viability3}
	\frac{\widehat{M}(t)}{M(t)} \approx \phi(t)=Ae^{\displaystyle -B\frac{e^{-\lambda t}-1+\lambda t}{\lambda}-C\frac{1-e^{-\lambda t}}{\lambda}},
\end{equation}
where $A\approx\displaystyle \frac{\widehat{M}(0)}{M(0)}=1$, $B=\alpha Q^*$, $C=\alpha\bar{Q}$. The values of $A$, $B$, $C$ and $\lambda$ for each concentration of the inhibiting agent can be estimated by fitting the curve $\phi$ to the experimental data.

\textbf{Remark.} The time scale on Figures \ref{Fig_Conv}, \ref{SlowLineApprox} and \ref{SlowNonlinApprox} is not specified. These figures represent qualitatively the behavior of the depicted functions over time. Note that the conclusions derived from these figures are of such nature that they are independent of the time scale. The time axis on Figures \ref{Fig_Conv}, \ref{SlowLineApprox} and \ref{SlowNonlinApprox} should not be confused with the time axis on Figure \ref{AveOverTime} as well as figures in the next sections, where the time is in hours and the range is determined by the measurements times in the experiments.

\section{Deriving approximation for the cell viability from the experimental data} \label{Section_5}

\subsection{Fitting curves of the form \eqref{eqn_cell_viability2} to the experimental data}

\begin{figure}[ht]
\includegraphics[trim=0cm 0cm 2cm 0cm,clip,width=0.5\textwidth]{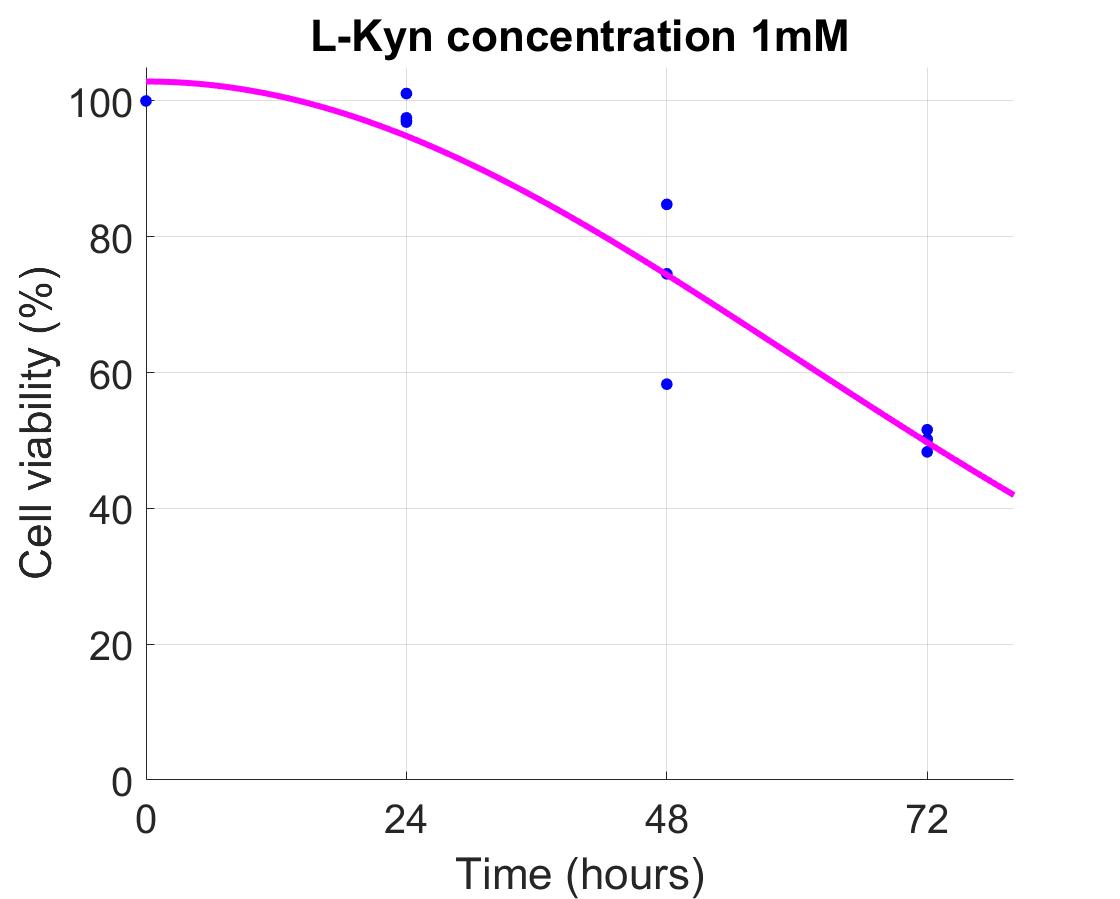}
\includegraphics[trim=0cm 0cm 2cm 0cm,clip,width=0.5\textwidth]{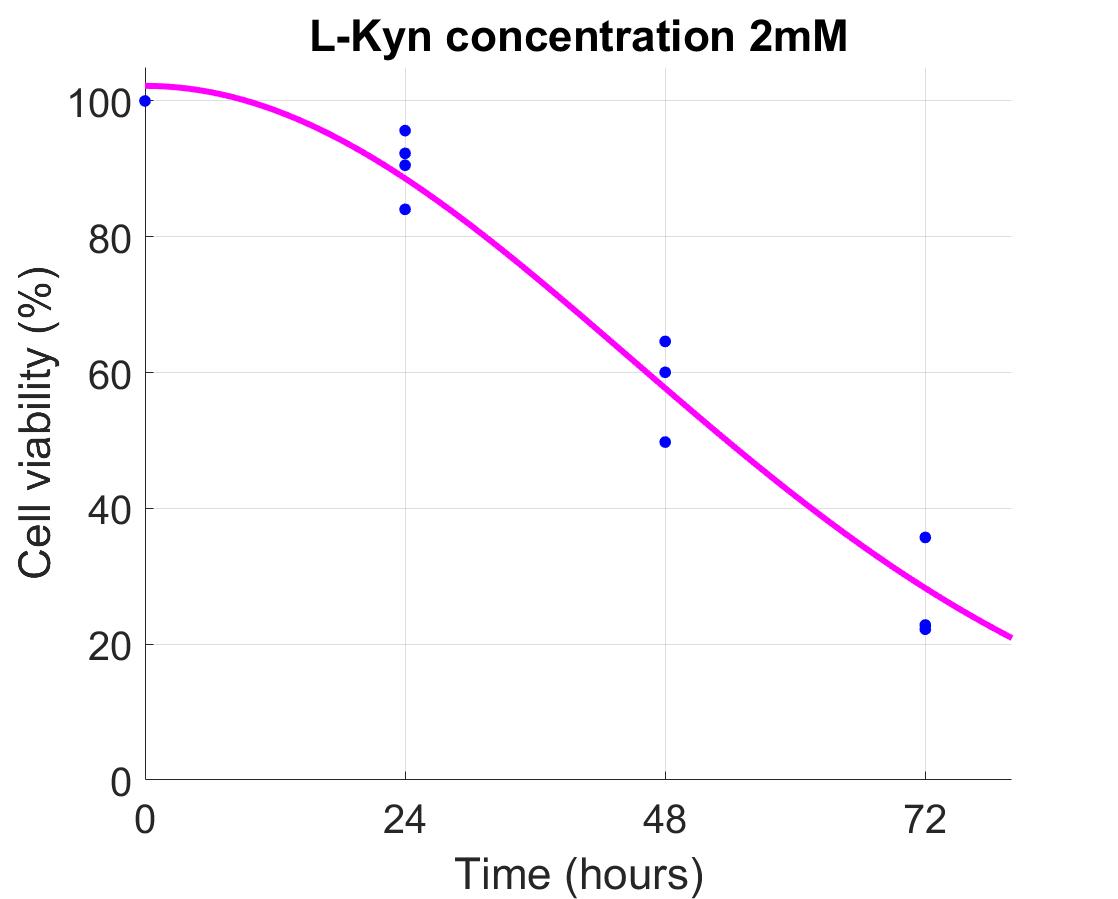}\\[12pt]
\includegraphics[trim=0cm 0cm 2cm 0cm,clip,width=0.5\textwidth]{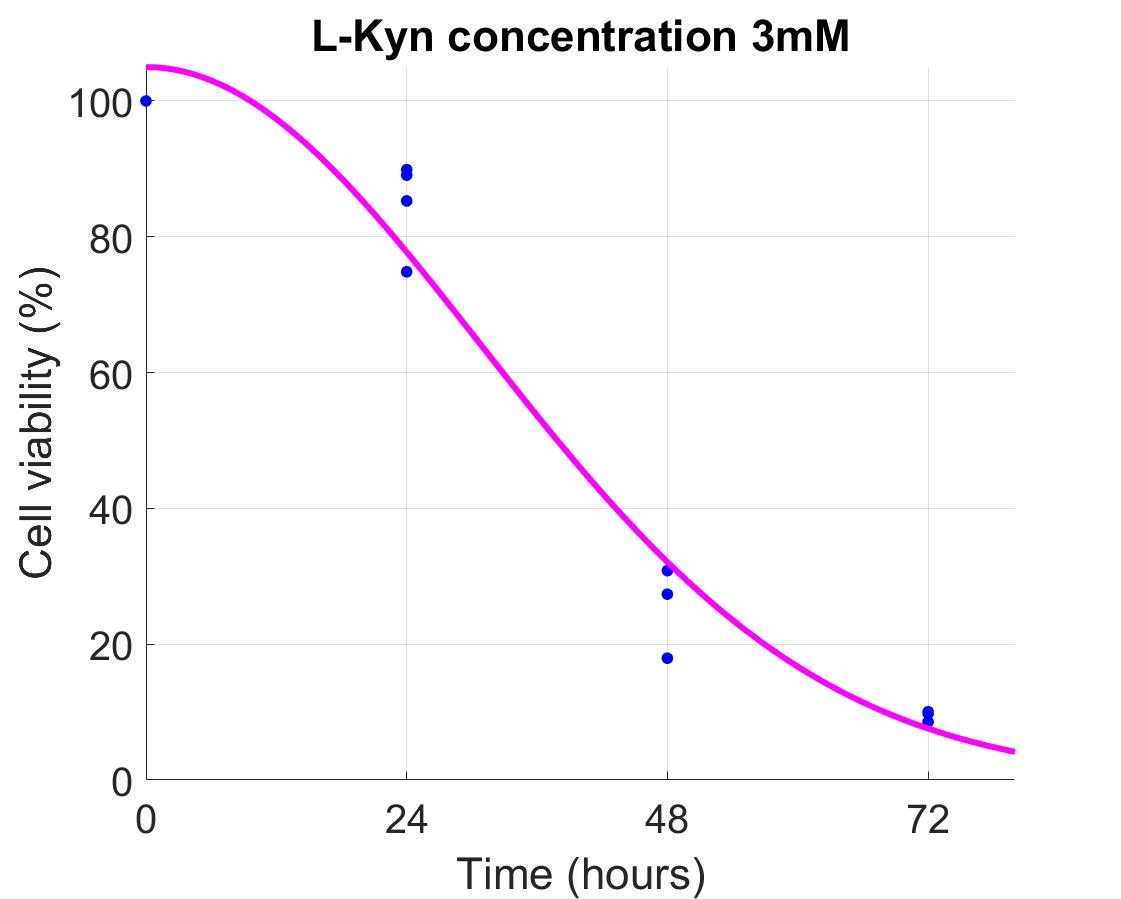}
\includegraphics[trim=0cm 0cm 2cm 0cm,clip,width=0.5\textwidth]{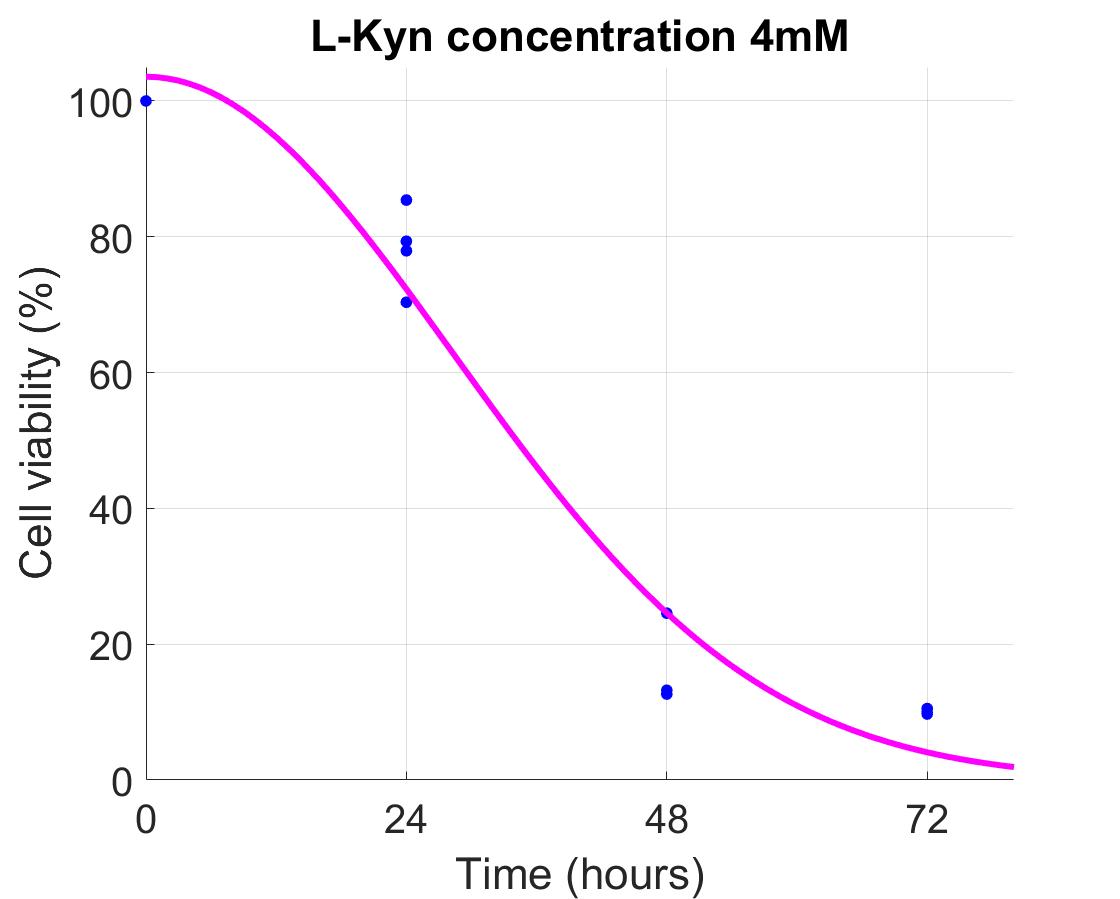}
\caption{Graphs of the functions of the form \eqref{phi4} fitted to the data at different concentrations. The blue dots represent the data points.}\label{fitting4}
\end{figure}

We consider the data discussed in Section 2, which was obtained via the crystal violet assay protocol using L-Kyn as an inhibiting agent. For any fixed concentrations, we use the least squares method to derive the best fitting curve of the form \eqref{eqn_cell_viability3}, where at $t=0$ we consider the square of $\phi(0)-1=A-1$. The optimization was performed using the Matlab function {\tt fminsearch} with starting points from a dense mesh covering the feasible domain of the parameter vector $(A,B,C,\lambda)$. The numerical procedure discovered multiple local equilibria with very similar values of the objective function, but very different values of the parameters. Further, all runs returned small values of $\lambda$ (not exceeding $10^{-3}$), which explains the ill-conditioning of the optimization problem. For small $\lambda t$, the first fraction in the exponent in \eqref{eqn_cell_viability3} is approximately equal to $\frac{1}{2}\lambda t^2$. Hence, $B$ and $\lambda$ each of which cannot be reliably estimated, e.g. increase in $C$ can be compensated by a decrease of $\lambda$, returning the same or very similar value of the objective function. This is resolved by rewriting the function $\phi$ in the form
\begin{equation}  \label{phi4}
	\phi(t)=Ae^{\displaystyle -D\frac{e^{-\lambda t}-1+\lambda t}{\lambda^2}-C\frac{1-e^{-\lambda t}}{\lambda}},
\end{equation}
where $D=\lambda B$.
This stabilises the optimization problem and we obtain a unique equilibrium for every concentration. The values of $A$, $D$, and $\lambda$ are given in Table 1, while the graphs of $\phi$ are presented in Figure \ref{fitting4}.

\begin{table}[ht]

\caption{Estimated values of the parameters $A$, $D$, $C$ and $\lambda$ in \eqref{phi4}}\vspace{12pt}

\centering
\begin{tabular}{|c|c|c|c|c|c|}\hline &&&&&\\[-8pt]
Concentration&$A$&$D$&$C$&$\lambda$&$R^2$\\[-8pt]&&&&&\\\hline &&&&&\\[-8pt]
1mM &1.02866 & $2.82646\times 10^{-4}$ & $1.47370\times 10^{-14}$ & $3.19765\times 10^{-4}$&0.9665 \\
2mM &1.02222 & $4.97166\times 10^{-4}$ & $3.22590\times 10^{-14}$ & $8.55739\times 10^{-5}$& 0.9794\\
3mM &1.04992 & $1.05762\times 10^{-3}$ & $1.40822\times 10^{-15}$ & $1.75403\times 10^{-3}$& 0.9710\\
4mM & 1.03818 & $1.27147\times 10^{-3}$ & $5.91781\times 10^{-16}$ & $1.55132\times 10^{-3}$& 0.9691 \\[-8pt]&&&&&\\\hline
\end{tabular}
\end{table}

Let us recall that the statistic $R^2$ measures the goodness of fit of a model. It represents the fraction of variation explained by the model. Therefore, it varies between 0 and 1. Values over 0.95, as the values obtained in Table 1, are considered to indicate a very good fit of the model to the data.

\subsection{Analysis and interpretation}

The main driver of the dynamics of the curves in Figure \ref{fitting4} is the term the parameter $D=\alpha\lambda Q^*$. Considering the mathematical definition and the biological meaning of $Q^*$ and $\lambda$, they should both increase with any increase of the concentration of the inhibiting agent. Hence, one expects that $D$ to increase too. Indeed, such increase of $D$ can be observed clearly in Table 1.

The parameter $C=\alpha\bar{Q}$ is expected to be more or less the same for all experiments since the initial states are similar. The obtained values of $C$ are in a short-range near zero,  from $10^{-14}$ to $10^{-16}$. These small values of $C$ imply that the initial states of the system are very close to the endpoint at $Q=0$ of the slow manifold. We recall that in the implemented experiments, the cells were allowed to settle under optimal conditions for 24 hours. One may expect that during this period large fraction of the receptors CXCR4 were occupied by the CXCL12 molecules. A small value of $C$, and respectively of $\bar{Q}$, support this hypothesis.

Regarding the estimated values of $\lambda$, firstly let us note that they are small.
Using $\bar{Q}\approx 0$, we have
$$
Q(t)\approx(1-e^{-\lambda t})Q^*.
$$
The small values of $\lambda$ indicate that the observed impact on cell viability occurs while $Q(t)$ is still relatively far from its equilibrium. As an  example, for the concentration of 3mM, $Q(72)\approx 0.118 Q^*$, that is the level of inhibition is about 12\% of its maximum for the given concentration.

\begin{figure}[ht]
\includegraphics[trim=0cm 0cm 2cm 0cm,clip,width=0.5\textwidth]{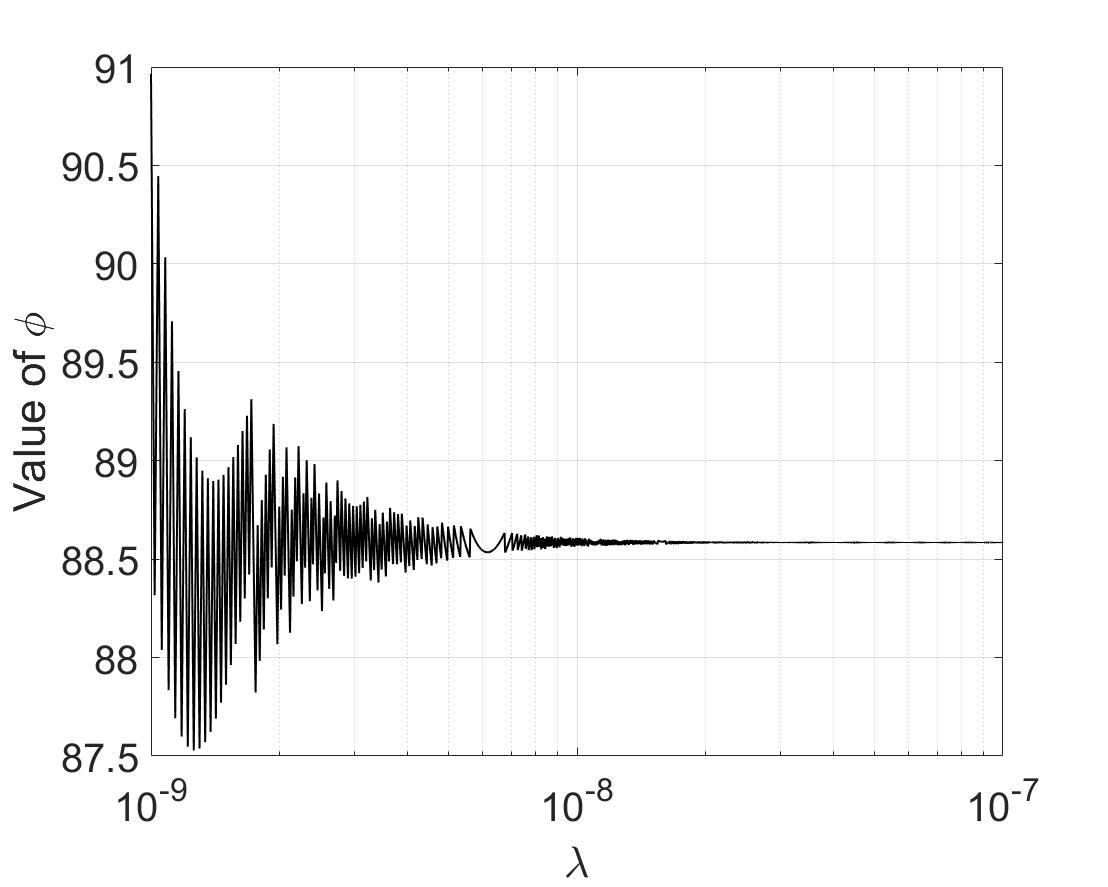}
\includegraphics[trim=0cm 0cm 2cm 0cm,clip,width=0.5\textwidth]{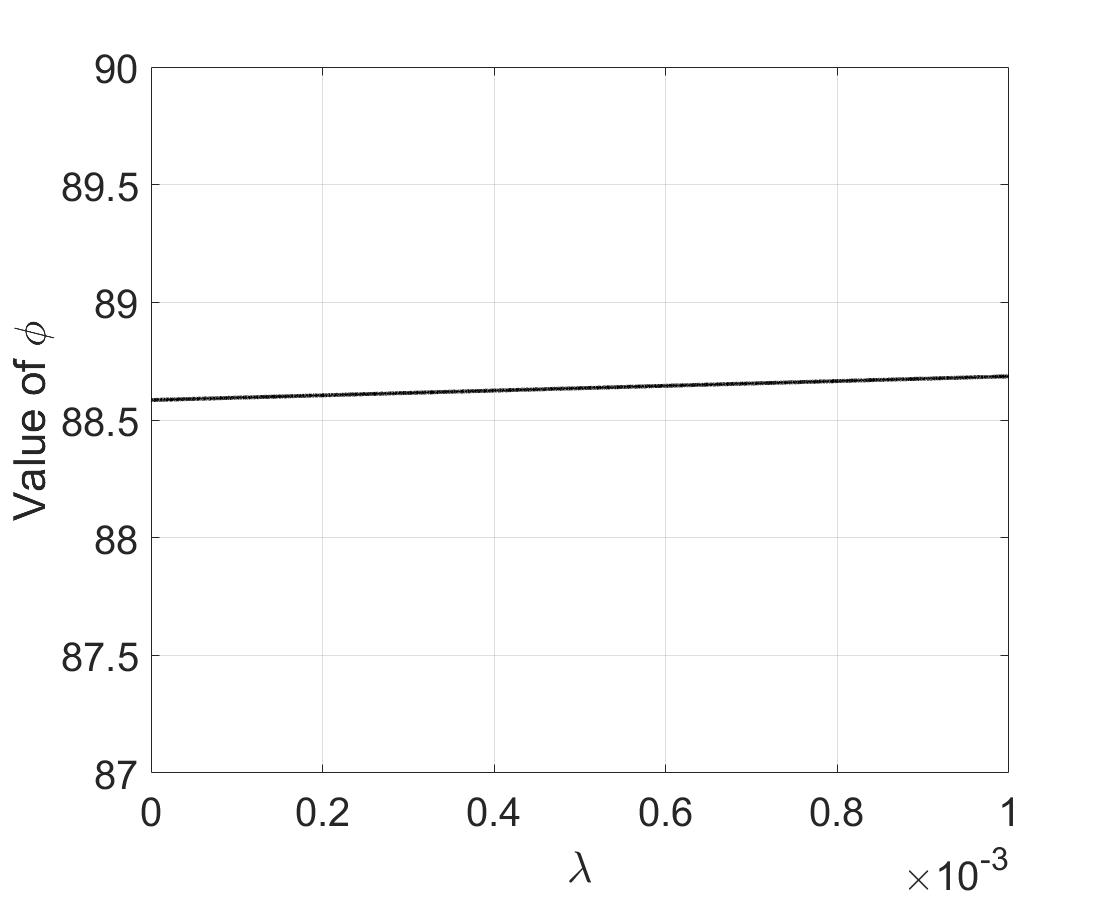}\\
(a)\hspace{0.5\textwidth} (b)
\caption{Graphs of $\phi$ as a the function of $\lambda$ with the other parameters as for concentration 2mM in Table 1 and $t=24$:
(a) $\lambda< 10^{-7}$ ($\log$ scale for $\lambda$, (b) $\lambda\!\in\![10^{-7},10^{-3}]$}\label{testingSmallL}
\end{figure}

Secondly, we can observe that in this range of $\lambda$ of about $10^{-3}$ and less, the function $\phi$ depends little on $\lambda$. As an example, we have on Figure \ref{testingSmallL} (b) the graph of $\phi$ as a function of $\lambda$ with the rest of the parameters being  for the concentration 2mM and $t=24$. There is indeed a very small gradient, which implies that the estimates of $\lambda$ are sensitive to small variations of the data. A general decreasing trend in the values of $\lambda$ can be noticed in Table 1, but it is not as clear and well pronounced as in the case of $D$. The mentioned sensitivity is a possible reason. From Figure \ref{testingSmallL}  (a),  we observe that for even smaller values of $\lambda$ ($\lambda<10^{-7}$) the accuracy of the computation of $\phi$ is completely lost due to roundoff errors.

\subsection{Reducing the number of parameters}

Theoretically, the function $\phi$ depends on four parameters. However, we established little dependence on $\lambda$. More precise computation shows that, we have
\begin{equation}\label{approx3}
\frac{e^{-\lambda t}-1+\lambda t}{\lambda^2}\approx \frac{1}{2}t^2,
\end{equation}
where for $\lambda<0.002$ and $t\leq 72$, the relative error of approximation is bounded above by
\begin{equation}\label{estLambda}
\frac{1}{3}\lambda t<4.8\%.
\end{equation}
Hence, we can apply the approximation formula in \eqref{approx3}. Similarly, we have
$$
\frac{1-e^{-\lambda t}}{\lambda}\approx t.
$$
However, due to the small value of $C$, the term containing this fraction makes a negligible contribution to the value of $\phi$. Therefore, from a mathematical point of view, the obtained functions $\phi$ can be nearly as well represented through functions from the two-parameter family
\begin{equation}  \label{psi}
	\psi(t)=Ae^{\displaystyle -\frac{1}{2}Dt^2}.
\end{equation}
To validate this statement we repeated the fitting process to the data using the functions $\psi$ as given in \eqref{psi}. The estimated values of $A$ and $D$ are given in Table 2. The values of $D$ are very similar to the values of $D$ in Table 1 with at least the first two significant figures being the same. The values of $A$ are the same correct to four significant figures.  The value of $R^2$ has a very small increase affecting the fourth or fifth digit only. The graphs of the functions $\psi$ with parameters from Table 2 are visually indistinguishable from the graphs of the functions $\phi$ on Figure ~\ref{fitting4}. Hence, these need not be presented.

\begin{table}[ht]
\caption{Estimated values of the parameters $A$ and $D$ in \eqref{psi}}\vspace{12pt}

\centering
\begin{tabular}{|c|c|c|c|}\hline &&&\\[-8pt]
Concentration&$A$&$D$&$R^2$\\[-8pt]&&&\\\hline &&&\\[-8pt]
1mM &1.02841 & $2.80609\times 10^{-4}$ & 0.9665 \\
2mM &1.02211 & $4.96267\times 10^{-4}$ & 0.9794\\
3mM &1.04874 & $1.02984\times 10^{-3}$ & 0.9716\\
4mM &1.03746 & $1.24571\times 10^{-3}$ & 0.9695 \\[-8pt]&&&\\\hline
\end{tabular}
\end{table}

The form \eqref{psi} takes into account that the parameters $C$ and $\lambda$ in \eqref{phi4} are confined into a small neighborhood of zero, so that one can get nearly as good approximation using a two parameter family functions as given in \eqref{psi}. The similarity of the approximations by $\phi$ and $\psi$ highlights
\begin{itemize}
\item the robustness of the estimation of these two parameters;
\item identifying the term $Dt^2$ as a primary driver of the dynamics with respect to time (due to this quadratic term we obtain the distinct sigmoidal shape of the curves in Figure \ref{fitting4});
\item the parameter $D$ captures the response to change of the concentration of the inhibitor.
\end{itemize}

\textbf{Remarks.}\begin{itemize}\item[1.] The parameter $D=\alpha \lambda Q^*$ is composite and depends on all parameters of the model. While $\lambda$ is removed from $\phi$ as we move from \eqref{phi4} to \eqref{psi}, $\lambda$ is not removed from the cell viability function \eqref{eqn_cell_viability2}. It is still represented through $D$. Further, the form \eqref{psi} shows that, while it might be difficult to estimate reliably $Q^*$ and $\lambda$ individually, a reliable estimation of the product comprising $D$ can be computed.
\item[2.]The fact that we can reduce the number of parameters of $\phi$ to two by letting $\lambda\to 0$ in the first fraction of the exponent and $C=0$ is only valid for the considered set of data and it is not a general property of the model. In a different experiment, where a different set of data is obtained, the situation may be different. For example, $\lambda$ is small only relative to the considered time interval $[0,72]$ so that \eqref{estLambda} holds. If experiments are conducted over a longer period of time, \eqref{approx3} would not be appropriate to use.
\item[3.]The parameter $A$ was introduced in \eqref{phi4} to account for possible variation in the initial states of the populations. The results in Table 1 and in Table 2 show that $A$ is persistently larger than 1 by 2\% to 5\%. This can be considered as an indication that $A$ is capturing some causal relationship not yet accounted for in the model. Every model is derived through some simplifying assumptions. In that sense, a model is never completely accurate. The mentioned overestimation of $A$ is rather small to affect the quality of the approximation, where decay due to the exponent in \eqref{psi} determines the primary dynamics. It is nevertheless an issue that can be investigated in future research theoretically and experimentally.
\end{itemize}

\section{Calculating $IC_{50}$} \label{Section_6}

As mentioned, $IC_{50}$ is a commonly used for characterising the inhibiting properties of an agent and as a benchmark for comparison with other agents. In more detail, given time $t$, $IC_{50}(t)$ is the concentration that reduces the viability of the cells by 50\% at time $t$. From the graphs in Figure \ref{fitting4}, given the respective concentration one can read the time for which this concentration provides a 50\% reduction of cell viability. We have $IC_{50}(71.84)=1$, $IC_{50}(53.69)=2$, $IC(37.86)=3$, $IC_{50}(34.20)=4$. We derive in this section the graph of the $IC_(50)$ as a function of time, so that one read from it the value of $IC_{50}$ for any specific time. For that purpose, we represent the cell viability as a function $\varphi(c,t)$ of concentration $c$ (in mM) of the inhibiting agent and the time $t$ (in hours).

\begin{figure}[h!]
\centering\includegraphics[scale=0.22]{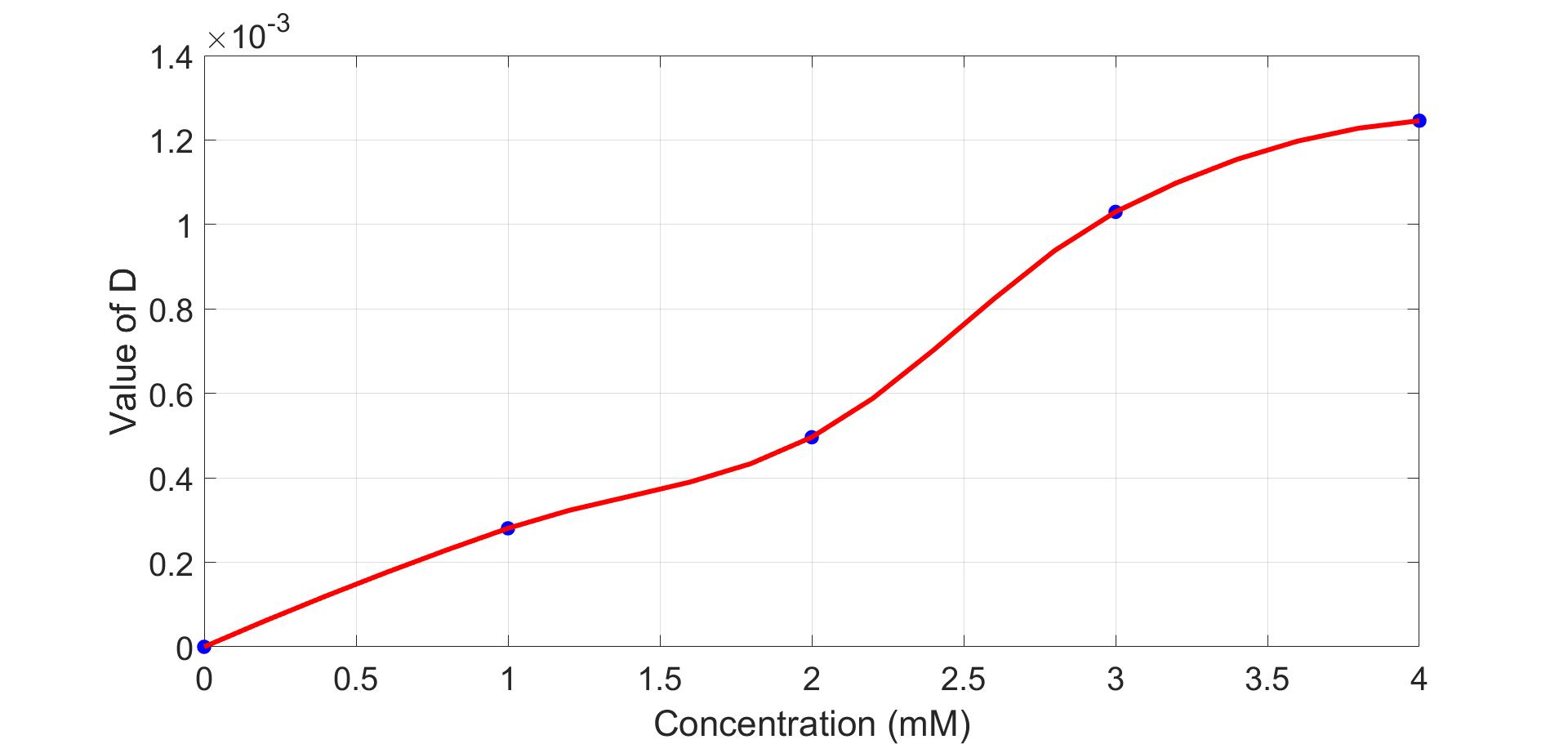}
\caption{Graphs of the values of $D$ for $c=0,1,2,3,4$ (blue dots) and the interpolating function for $c\in[0,4]$ (red solid line)}\label{Fig_D}\vspace{12pt}
\end{figure}

As shown in Section 5, the impact of the concentration of the inhibitor is captured through the change of the parameter $D$ in \eqref{psi}. Let us consider $D$ as a function of $c$, that is $D=D(c)$. The second column of Table 2 gives the values of $D(1)$, $D(2)$, $D(3)$ and $D(4)$. By default $D(0)=1$. Since we do not have in any explicit form how $D$ depends on $c$, we would not attempt a curve fitting, but rather use an interpolation of the existing data. One can reasonably expect that the function is $D$ is smooth. Hence, we apply interpolation via cubic convolution, which provides for a smooth function (unlike e.g. the linear interpolation), while staying very close to the data as the weighted average of the nearest 4 points (3 points for the boundary intervals) is used \cite{Keys}. The graph of the data points and interpolating function $D(c)$ for $c\in[0,4]$ is shown in Figure \ref{Fig_D}.

\begin{figure}[t!]
\centering\includegraphics[scale=0.22]{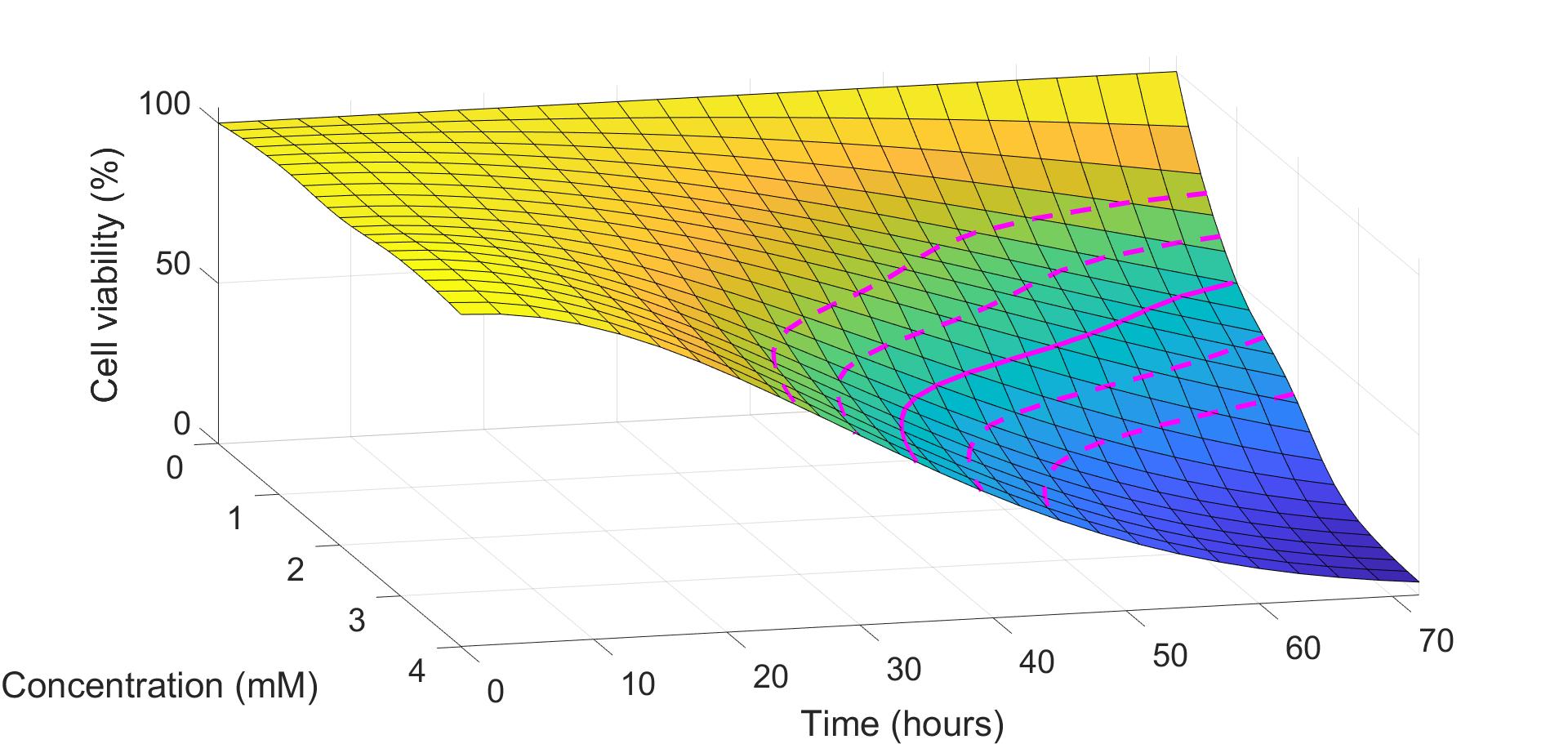}
\caption{Surface graph of the cell viability function $\varphi$ defined in \eqref{varphi}. Solid line: level curve at 50\%. Dashed lines: level curves at 30\%, 40\%, 60\% and 70\%.}\label{viability}

\centering\includegraphics[scale=0.22]{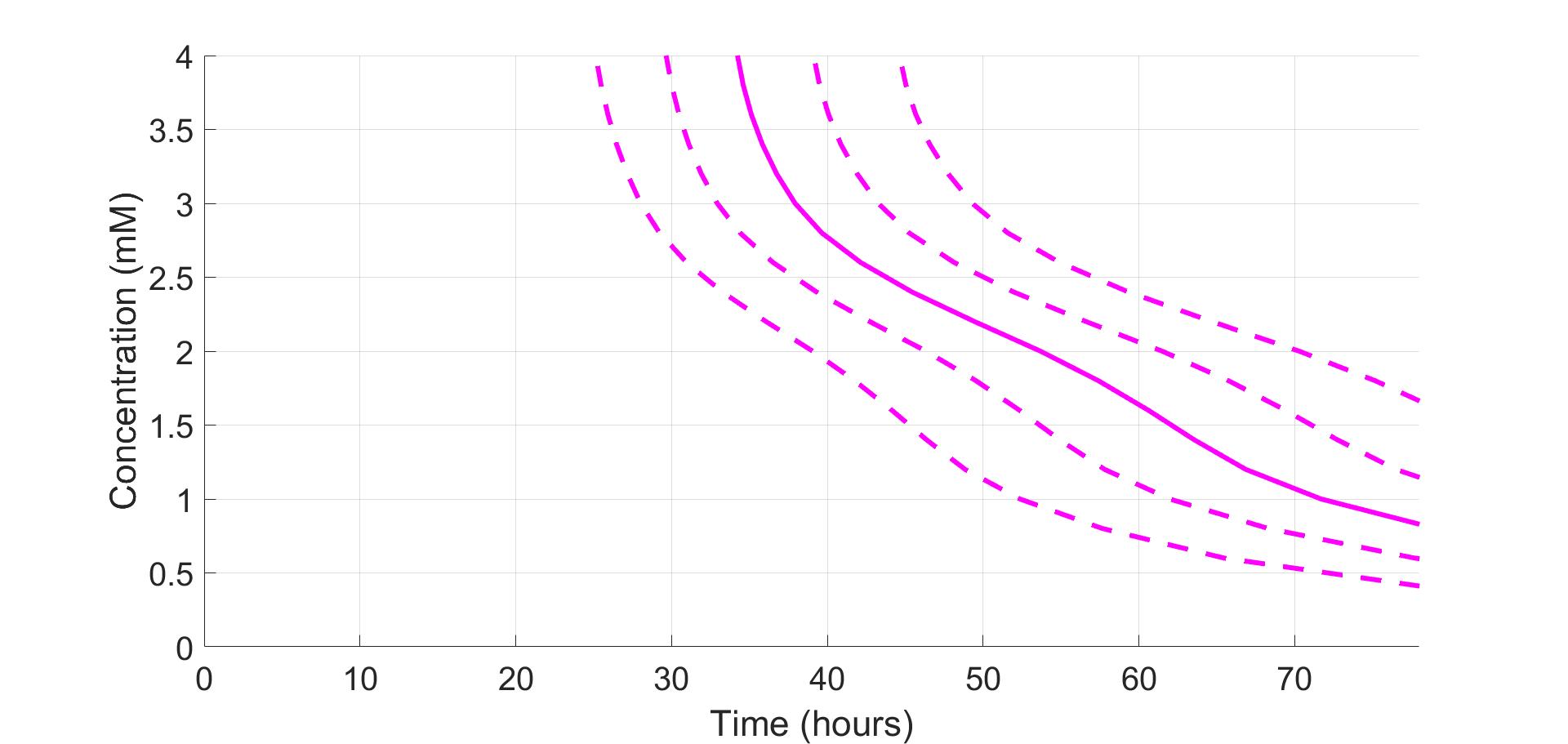}
\caption{Solid line: $IC_{50}(t)$ as a function of time. Dashed lines (left to right): $IC_{30}(t)$, $IC_{40}(t)$, $IC_{60}(t)$ and $IC_{70}(t)$ as functions of time. }\label{IC50curve}
\end{figure}

In a similar way, we obtain the function $A(c)$ for $c\in[0,4]$. We omit the details since $A$ exhibits little change for different concentrations with respectively little impact on the cell viability.

Thus we obtain the cell viability function
\begin{equation}\label{varphi}
\varphi(c,t)=A(c)e^{\displaystyle-\frac{1}{2}D(c)t^2}, \ c\in[0,4],\ t\in [0,72].
\end{equation}

The graph of the viability function $\varphi$ is given on Figure \ref{viability}. The magenta solid line on the surface represents the intersection with the horizontal plane where cell viability $=50\%$. When plotted on concentration vs. time axes, we obtain the graph of the $IC_{50}$ as a function of time, see solid line on Figure \ref{IC50curve}. We can read from the graph $IC_{50}$ for any given time. For example, we have $IC_{50}(48)=2.2725$, $IC_{50}(60)=1.638$. Further, since we have the cell viability in the explicit form \eqref{varphi}, we can construct a level line at any cell viability level. The dashed lines on Figures \ref{viability} represent level lines for cell viability of 30\%, 40\%, 60\% and 70\%. Hence, we can obtain not only the half maximum inhibitory concentration ($IC_{50}$), but also the inhibitory concentration for any required level of cell viability at any specified time. The dashed lines represent the inhibitory concentrations for cell viability of 30\%, 40\%, 60\% and 70\% as functions of time.

\section{Conclusions}

The paper represents a study of the inhibition of cancer cell viability via blocking a signalling pathway essential for the adhesion and proliferation of the cells. The method of analysis is based on the derivation of a mathematical model representing the inhibition mechanism and the activation-inhibition dynamics. The experimental results are assimilated and interpreted through the model. Thus, variations in the data which may be obscuring general trends are "filtered out". The method is exemplified on blocking the CXCR4/CXCL12 axis in melanoma cells. The experimental data is obtained by using L-Kynurenine as an inhibiting agent. The five-step protocol given in the introduction is implemented and provides a variety of output for the inhibition mechanism and the experiment as reported in Section~\ref{Section_5}. The $IC_{50}$ curve in Section~\ref{Section_6} is relevant to any further consideration of the feasibility of treatment via this inhibitor.

Using the same integrated approach of mathematical modelling and experimental work one can consider other inhibitors. Specifically, this team intends to study the impact of the CTCE-9908 agent (e.g., \cite{porvasnik2009effects}) on the cell viability of melanoma cells. However, the method is open to a wider spectrum of applications. Further attention will be given to unveiling any unaccounted yet causal relationships in the model, e.g. the reason for the persistent overestimation of the parameter $A$. Certainly, such work will improve the accuracy of the model and the reliability of the produced results.

\section*{Acknowledgements}
The authors would like to thank Charlise Basson, MSc student, Department of Physiology, University of Pretoria, for her contribution in procuring the data and creating the CXCR4 cancer cell and inhibitor image in Figure 1.

The research was supported by the DST/NRF SARChI Chair on Mathematical Models and Methods in Bioengineering and Biosciences at the University of Pretoria.

A E Phiri acknowledges the support of IMU through a Breakout Graduate Fellowship Grant.

\section*{Declaration of interests: None}


\bibliographystyle{pnas-new}
\bibliography{drug.bib}

\begin{thebibliography}{10}

\bibitem{vera2021mathematical}
Vera J, et~al. (2021) Mathematical modelling in biomedicine: A primer for the
  curious and the skeptic.
\newblock {\em International journal of molecular sciences} 22(2):547.

\bibitem{mendoza2012mathematical}
Mendoza-Juez B, Mart{\'\i}nez-Gonz{\'a}lez A, Calvo GF, P{\'e}rez-Garc{\'\i}a
  VM (2012) A mathematical model for the glucose-lactate metabolism of in vitro
  cancer cells.
\newblock {\em Bulletin of mathematical biology} 74(5):1125--1142.

\bibitem{benzekry2014classical}
Benzekry S, et~al. (2014) Classical mathematical models for description and
  prediction of experimental tumor growth.
\newblock {\em PLoS computational biology} 10(8):e1003800.

\bibitem{lima2016selection}
Lima E, Oden J, Hormuth D, Yankeelov T, Almeida R (2016) Selection,
  calibration, and validation of models of tumor growth.
\newblock {\em Mathematical Models and Methods in Applied Sciences}
  26(12):2341--2368.

\bibitem{hixson1931dependence}
Hixson A, Crowell J (1931) Dependence of reaction velocity upon surface and
  agitation.
\newblock {\em Industrial \& Engineering Chemistry} 23(8):923--931.

\bibitem{korsmeyer1983mechanisms}
Korsmeyer RW, Gurny R, Doelker E, Buri P, Peppas NA (1983) Mechanisms of solute
  release from porous hydrophilic polymers.
\newblock {\em International journal of pharmaceutics} 15(1):25--35.

\bibitem{peppas2014mathematical}
Peppas NA, Narasimhan B (2014) Mathematical models in drug delivery: How
  modeling has shaped the way we design new drug delivery systems.
\newblock {\em Journal of Controlled Release} 190:75--81.

\bibitem{CVA2016}
Feoktistova M, Geserick P, Leverkus M (2016) Crystal violet assay for
  determining viability of cultured cells.
\newblock {\em Cold Spring Harbor Protocols, doi:10.1101/pdb.prot087379}.

\bibitem{haibe2013inconsistency}
Haibe-Kains B, et~al. (2013) Inconsistency in large pharmacogenomic studies.
\newblock {\em Nature} 504(7480):389--393.

\bibitem{geeleher2016consistency}
Geeleher P, Gamazon ER, Seoighe C, Cox NJ, Huang RS (2016) Consistency in large
  pharmacogenomic studies.
\newblock {\em Nature} 540(7631):E1--E2.

\bibitem{orimo2005stromal}
Orimo A, et~al. (2005) Stromal fibroblasts present in invasive human breast
  carcinomas promote tumor growth and angiogenesis through elevated
  sdf-1/cxcl12 secretion.
\newblock {\em Cell} 121(3):335--348.

\bibitem{cardones2003cxcr4}
Cardones AR, Murakami T, Hwang ST (2003) Cxcr4 enhances adhesion of b16 tumor
  cells to endothelial cells in vitro and in vivo via $\beta$1 integrin.
\newblock {\em Cancer research} 63(20):6751--6757.

\bibitem{wong2008translating}
Wong D, Korz W (2008) Translating an antagonist of chemokine receptor cxcr4:
  from bench to bedside.
\newblock {\em Clinical Cancer Research} 14(24):7975--7980.

\bibitem{khinkis2003optimal}
Khinkis LA, Levasseur L, Faessel H, Greco WR (2003) Optimal design for
  estimating parameters of the 4-parameter hill model.
\newblock {\em Nonlinearity in biology, toxicology, medicine}
  1(3):15401420390249925.

\bibitem{Tabor}
Tabor M (1988) {\em Chaos and Integrability in Nonlinear Dynamics: An
  Introduction}.
\newblock (Wiley).

\bibitem{Wiggins}
Wiggins S (2003) {\em Introduction to Applied Nonlinear Dynamical Systems and
  Chaos}.
\newblock (Springer).

\bibitem{walczak2020effect}
Walczak K, et~al. (2020) Effect of tryptophan-derived ahr ligands, kynurenine,
  kynurenic acid and ficz, on proliferation, cell cycle regulation and cell
  death of melanoma cells—in vitro studies.
\newblock {\em International Journal of Molecular Sciences} 21(21):7946.

\bibitem{marszalek2021kynurenine}
Marszalek-Grabska M, et~al. (2021) Kynurenine emerges from the shadows--current
  knowledge on its fate and function.
\newblock {\em Pharmacology \& Therapeutics} 225:107845.

\bibitem{Keys}
Keys RG (1982) Cubic convolution interpolation for digital image processing.
\newblock {\em IEEE Transactions on Acoustics Speech and Signal Processing}
  29(6):1153 -- 1160.

\bibitem{porvasnik2009effects}
Porvasnik S, et~al. (2009) Effects of cxcr4 antagonist ctce-9908 on prostate
  tumor growth.
\newblock {\em The Prostate} 69(13):1460--1469.

\bibitem{layek2015introduction}
Layek G (2015) {\em An introduction to dynamical systems and chaos}.
\newblock (Springer) Vol.{} 449.

\bibitem{dumortier2006qualitative}
Dumortier F, Llibre J, Art{\'e}s JC (2006) {\em Qualitative theory of planar
  differential systems}.
\newblock (Springer).

\end{thebibliography}

\section*{Appendix}

Before, we present the proof of Theorem~\ref{Thm_eqm}, we recall the Bendixon-Dulac  criterion (e.g., \cite{layek2015introduction}) concerning the non-existence of a periodic solution for a planar system. For a planar system $\dot{x}=f(x,y)$ and $ \dot{y} = g(x,y)$ with a vector field $F=(f,g)$ if $div(F) = \partial f/\partial x + \partial g/\partial y \not= 0$ for any region $D$ on a simple connected set (a region of the plane without gaps/holes) then there is no non-constant periodic solution in $D$.

{\bf Proof of Theorem~\ref{Thm_eqm}}
{\bf Proof of (i). } The equilibria of (\ref{eqP2})--(\ref{eqQ2}) are solutions of the simultaneous equations
\begin{eqnarray}
&&k_1(R_0-P-Q)(L_0-P)=k_{-1}P\label{eqPequi},\\
&&k_2(R_0-P-Q)(X_0-Q)=k_{-2}Q\label{eqQequi}.
\end{eqnarray}
Dividing the left hand sides and the right hand sides of the equations \eqref{eqPequi} and \eqref{eqQequi} yields
\begin{equation}\label{ratio}
\frac{k_1(L_0-P)}{k_2(X_0-Q)}=\frac{k_{-1}P}{k_{-2}Q}.
\end{equation}
Solving for $Q$ we obtain
\begin{equation}\label{eq1equi}
Q=\frac{X_0P}{P+\frac{k_1k_{-2}}{k_2k_{-1}}(L_0-P)}.
\end{equation}
Expressing $Q$ from equation (\ref{eqPequi}) we obtain
\begin{equation}\label{eq2equi}
Q=R_0-P-\frac{k_{-1}P}{k_1(L_0-P)}.
\end{equation}
The system \eqref{eqPequi}--\eqref{eqQequi} is equivalent to the system of equations \eqref{eq1equi}--\eqref{eq2equi}, where each equation gives $Q$ as a function of $P$.
Equation (\ref{eq1equi}) defines a continuous increasing function of $P$ on the interval $[0,L_0]$.  Considered on the interval $[0,L_0]$ its graph connects the points $(0,0)$ and $(L_0,X_0)$.
The function of $P$ given in (\ref{eq2equi}) is continuous and decreasing on the interval $[0,L_0)$ from $R_0$ to $-\infty$.
Then we can conclude that the graphs of the function in (\ref{eq1equi}) and (\ref{eq2equi}) intersect exactly once at a point $(P^*,Q^*)$, which satisfies $0\le P^* \le L_0$ and $0 \le Q^* \le X_0$.  Therefore, $(P^*,Q^*)$ is a unique solution of (\ref{eqPequi})--(\ref{eqQequi}) in $D$ or, equivalently, a unique equilibrium of (\ref{eqP2})--(\ref{eqQ2}).

{\bf Proof of (ii). }   The Jacobian $J$ of the vector field $F$ of the system in (\ref{eqP2})--(\ref{eqQ2})
can be written as
$$ J= \left( \begin{array}{cc}
k_1(\alpha+\beta)-k_{-1} & k_1\beta \\
k_2\gamma & k_2(\alpha+\gamma)-k_{-2}
\end{array} \right),$$
where $\alpha := P+Q - R_0$, $\beta := P-L_0$ and $\gamma := Q-X_0$.  Observe $\alpha,\beta,\gamma<0$ when $(P,Q)$ is contained in the interior of $D$ and  $\alpha,\beta,\gamma\le 0$ when $(P,Q)$ is contained in $D$.

We do not evaluate the point $(P^*,Q^*)$ to avoid dealing with tedious expressions, but instead show that it is a hyperbolic sink by finding that the Jacobian evaluated at $(P^*,Q^*)$ has to have eigenvalues $\lambda_1$ and $\lambda_2$, only to the left of the imaginary axis of the complex plane. To show both $\lambda_1$ and $\lambda_2$ lie only to the left of the imaginary axis of the complex plane, it is sufficient to show that the product $\lambda_1\lambda_2 = \mbox{det}(J)_{|(P^*,Q^*)}$ is positive while $\lambda_1+\lambda_2 = \mbox{Trace}(J)_{|(P^*,Q^*)}$ is negative.
The determinant of $J$, $\mbox{det}(J)$ simplifies to

\begin{eqnarray*}\mbox{det}(J)
	&=& k_1k_2(\alpha+\beta)(\alpha+\gamma) - k_{1}k_{-2}(\alpha + \beta) -  k_{-1}k_{2}(\alpha + \gamma) + k_1k_2\beta\gamma + k_{-1}k_{-2} \\
	&=& k_1k_2\bigg((\alpha+\beta)(\alpha+\gamma)-\beta\gamma)\bigg)- k_{1}k_{-2}(\alpha + \beta) -  k_{-1}k_{2}(\alpha + \gamma) + k_{-1}k_{-2} \\
	& =& k_1k_2\alpha(\alpha+\beta+\gamma) - k_{1}k_{-2}(\alpha + \beta) -  k_{-1}k_{2}  (\alpha + \gamma) + k_{-1}k_{-2}.\end{eqnarray*}

Since $\alpha,\beta,\gamma<0$ in the interior of $D$,
and $(P^*,Q^*)$ lies on the boundary of $D$,
the terms $k_1k_2\alpha(\alpha+\beta+\gamma)$, $- k_{1}k_{-2}(\alpha + \beta)$ and $-  k_{-1}k_{2}(\alpha + \gamma)$ are both positive, and thus $\mbox{det}(J)>0$.

Next, in the same vein, since $\alpha,\beta,\gamma \le 0$ on $D$, we have $\mbox{trace}(J) = \mbox{div}(F)=(k_1(\alpha+\beta)-k_{-1}) + k_2(\alpha+\gamma)-k_{-2})<0$ on all of $D$. Thus $(P^*,Q^*)$ is a hyperbolic sink, and by the Hartman-Grobman theorem it is also locally asymptotically stable. Further, the product of the non-diagonal entries of $J$ is positive, which implies that the eigenvalue of $J$ are real and distinct. Hence, the equilibrium is a stable proper node.

{\bf Proof of (iii). } We know that when the vector field $F$ on an open interval $U\subset \mathbb{R}^2$ is of class $C^r$, where $1 \le r < \infty$, for any $x \in U$, there exists an interval $I_x \subset \mathbb{R}$ on which the solution $\phi(0)$ so that $\phi(0)=x$ is unique (e.g., \cite[Theorem 1.1]{dumortier2006qualitative}).
Hence any initial value problem involving the system (\ref{eqP2})--(\ref{eqQ2}) has a unique local solution, and we can define the flow of the system. Let $\Phi(t,x)$ where $\Phi : \mathbb{R} \times \mathbb{R}^2 \to \mathbb{R}^2$ denote the flow of (\ref{eqP2}) --(\ref{eqQ2}).

{\bf Claim. } There is no solution with initial condition in $D$ that leaves $D$ except possibly at the point $(R_0,0)$, i.e., $\Phi(t,x) \in D$ for all $t\ge 0$ whenever $x \in
 D \setminus  (R_0,0)$.

{\bf Proof of Claim. } Due to the continuity of the flow and since $D$ is a connected subset of $\mathbb{R}^2$, if $\Phi(t_1,x) \in D$ and $\Phi(t_2,x) \notin D$ for some $t_1< t_2$ then there is a  $t_1 < \tau < t_2$ so that
$\Phi(t_1,x) \in \partial D$,
where $\partial D$ denotes the boundary of $D$. The boundary of $D$ consists of four line segments:
\begin{align*}
l_1&:=\{(P,Q)|P=0,~0\leq Q \leq R_0\},\\
l_2&:=\{(P,Q)|Q=0,~0\leq P\leq L_0\},\\
l_3&:=\{(P,Q)|P=L_0,~0\leq L_0\leq R_0\},\\
l_4&:=\{(P,Q)|P + Q= R_0\}.
\end{align*}

Consider $l_1$. Suppose for some $(P,Q) \in l_1$, there exists $\tau_{PQ}>0$ so that $\Phi(t,(P,Q)) \notin D$ for $0<t < \tau_{PQ}$. Then by the definition of the set $D$ it follows that $\frac{dP}{dt} \le 0$.
The vector field evaluated at any point $(P,Q) \in l_1$ is $F(P,Q) = (k_1(R_0-Q)L_0, k_2(R_0-Q)(X_0-Q) - k_{-2}Q )$. Since  $k_1(R_0-Q)L_0 >0$ for all $Q$ in the interval $[0,R_0)$,  $\frac{dP}{dt} > 0$. Hence $\Phi(t,(P,Q)) \in D$ for all $t< \tau_{P,Q}$ whenever $(P,Q) \in l_1 \setminus \{ (0,R_0) \}$.


Suppose for some $(P,Q) \in l_2$, there exists $\tau_{PQ}>0$ so that $\Phi(t,(P,Q)) \notin D$ for $0<t < \tau_{PQ}$. Then by the definition of the set $D$ it follows that $\frac{dQ}{dt} \le 0$.
For $(P,Q) \in l_2$, we have $F(P,Q) = (k_1(R_0-P)(L_0-P)-k_{-1}P, k_2(R_0-P)(X_0))$. Since we have assumed all constants to be non-zero and positive, we have $R_0 > L_0$. Hence  $k_2(R_0-P)X_0 >0$ for all $P$ in the interval $[0,L_0]$,  $\frac{dQ}{dt} > 0$.  Hence $\Phi(t,(P,Q)) \in D$ for all $t< \tau_{P,Q}$ whenever $(P,Q) \in l_2 \}$.

Suppose for some $(P,Q) \in l_3$, there exists $\tau_{PQ}>0$ so that $\Phi(t,(P,Q)) \notin D$ for $0<t < \tau_{PQ}$. Then by the definition of the set $D$ it follows that $\frac{dP}{dt} \ge 0$.

For $(P,Q) \in l_3$,  $F(P,Q) = -k_{-1}L_0, k_2(R_0-L_0-Q)(X_0-Q) - k_{-2} Q)$. Since $-k_{-1}L_0 < 0$, $\frac{dP}{dt} <0$, and hence  $\Phi(t,(P,Q)) \in D$ for all $t< \tau_{P,Q}$ whenever $(P,Q) \in l_3 \}$.

Suppose for some $(P,Q) \in l_3$, there exists $\tau_{PQ}>0$ so that $\Phi(t,(P,Q)) \notin D$ for $0<t < \tau_{PQ}$. Then by the definition of the set $D$ it follows that $\frac{dP}{dt} \ge 0$ and $\frac{dQ}{dt} \ge 0$. For $(P,Q) \in l_4$, $F(P,Q) = -k_{-1}P,  - k_{-2} Q)$, and since $k_{-1}P,k_{-2}Q$ are both negative on $l_4 \setminus \{ (0,R_0),  (L_0,0)\}$, and hence for each point $(P,Q) \in l_4\setminus \{ (0,R_0),  (L_0,0)\}$, there exists an $\tau_{PQ} > 0$ such that $\Phi(t,(P,Q)) \in D$.
This proves the claim.

It remains to be shown that there is no trajectory that leaves the set $D$ through the point $(0,R_0)$. When the vector field $F$ is $C^r$, then the flow $\Phi$ is also $C^r$ (\cite[Theorem 1.1]{dumortier2006qualitative}), and in particular, the map $f_\epsilon: x \mapsto \Phi(\epsilon,x)$ is a homeomorphism for every $\epsilon$.  Suppose that for some $y \in \partial D$, $\Phi(t,y) \notin D$ for every $0<t\le\epsilon_0$.  Now consider the map $f_{\epsilon_{0}}: x \mapsto \Phi(\epsilon_0,x)$. Since $f_{\epsilon_{0}}$ is a homeomorphism, if $f_{\epsilon_{0}}(y) \notin D$, then there is an open set $V$ containing $y$ so that $f_\epsilon(V)$ belongs to the complement of $D$.  Since the solutions that leave $D$, has to pass through the point $(0,R_0)$ in view of the above claim, this violates the uniqueness of solutions.

\textbf{Proof of iv}.  Suppose there exists a point $\bar{r}\in D$ that is not contained in the basin of the sink $(P^*,Q^*)$. This implies that the $\omega$-limit set of the trajectory passing through $\bar{r}$ is contained in $D$ by (iii). Owing to a result of a Poincare and Bendixon (e.g., \cite{dumortier2006qualitative}), for a planar flow, the $\omega$-limit set would either be an equilibrium solution or a periodic solution or a saddle-loop .  Since $div(F)<0$ on $D$, a periodic solution is ruled out by Bendixon-Dulac criterion, and since $(P^*,Q^*)$ is the unique equilibrium that is a sink, a saddle loop is also not possible. Thus, no such $\bar{r}$ exists, and hence $D$ is contained within the basin of the hyperbolic sink.   $\blacksquare$

\end{document}